\documentclass[runningheads]{llncs}

\newcommand{\ExtendedVersion}[1]{#1}\newcommand{\PaperVersion}[1]{}

\usepackage[T1]{fontenc}
\usepackage{amsmath, amssymb}					
\usepackage[english]{babel}
\usepackage{hyperref}
\usepackage{multirow}
\usepackage{latexsym}
\usepackage{listings}
\usepackage{booktabs}

\usepackage{graphicx}
\usepackage[dvipsnames,svgnames, table]{xcolor} 
\DeclareFontFamily{U}{mathx}{\hyphenchar\font45}
\DeclareFontShape{U}{mathx}{m}{n}{
      <5> <6> <7> <8> <9> <10>
      <10.95> <12> <14.4> <17.28> <20.74> <24.88>
      mathx10
      }{}
\DeclareSymbolFont{mathx}{U}{mathx}{m}{n}
\DeclareMathSymbol{\bigtimes}{1}{mathx}{"91}

\usepackage{etoolbox} 
\usepackage{orcidlink} 

\usepackage[linesnumbered, vlined, ruled]{algorithm2e}
	\SetAlCapHSkip{0px} 
	\SetNlSty{}{}{}     

\makeatletter
\patchcmd{\algocf@makecaption@ruled}{\hsize}{\textwidth}{}{} 
\patchcmd{\@algocf@start}{-1.5em}{0em}{}{} 
\makeatother

\makeatletter
\renewcommand{\paragraph}{%
  \@startsection{paragraph}{4}%
  {\z@}{1.5ex \@plus 1ex \@minus .2ex}{-1em}%
  {\normalfont\normalsize\bfseries}%
}
\makeatother

\newcounter{querycnt}
\newcommand{\QueryLabel}[2]{\noindent\small\textbf{Query~\refstepcounter{querycnt}\arabic{querycnt}.} #2 \label{#1}}

\definecolor{eclipseStrings}{RGB}{42,0.0,255}
\definecolor{eclipseKeywords}{RGB}{127,0,85}
\definecolor{darkgreen}{RGB}{0,128,21}
\colorlet{numb}{magenta!60!black}
\usepackage[normalem]{ulem} 
\makeatletter
\def\uwave{\bgroup \markoverwith{\lower3.5\p@\hbox{\sixly \textcolor{red}{\char58}}}\ULon}
\font\sixly=lasy6 
\makeatother

\usepackage{xargs}                      
\usepackage[colorinlistoftodos,prependcaption,textsize=tiny]{todonotes}
\newcommandx{\unsure}[2][1=]{\todo[linecolor=red,backgroundcolor=red!25,bordercolor=red,#1]{#2}}
\newcommandx{\change}[2][1=]{\todo[linecolor=blue,backgroundcolor=blue!25,bordercolor=blue,#1]{#2}}
\newcommandx{\info}[2][1=]{\todo[linecolor=OliveGreen,backgroundcolor=OliveGreen!25,bordercolor=OliveGreen,#1]{#2}}
\newcommandx{\improvement}[2][1=]{\todo[linecolor=Plum,backgroundcolor=Plum!25,bordercolor=Plum,#1]{#2}}
\newcommandx{\thiswillnotshow}[2][1=]{\todo[disable,#1]{#2}}
\newcommandx{\todoi}[2][1=]{\todo[inline,size=\small,linecolor=Plum,backgroundcolor=Periwinkle!25,#1]{#2}}
\newcommandx{\todoCite}[2][1=]{\todo[linecolor=Plum,backgroundcolor=Periwinkle!25,bordercolor=Plum,#1]{#2}}
\newcommandx{\todoDef}[2][1=]{\todo[linecolor=Plum,backgroundcolor=Periwinkle!25,bordercolor=Plum,#1]{Def.: #2}}

\makeatletter
\font\uwavefont=lasyb10 scaled 700
\def\spelling{\bgroup\markoverwith{\lower3.5\p@\hbox{\uwavefont\textcolor{Red}{\char58}}}\ULon}
\def\grammar{\bgroup\markoverwith{\lower3.5\p@\hbox{\uwavefont\textcolor{LimeGreen}{\char58}}}\ULon}
\def\phrasing{\bgroup\markoverwith{\lower3.5\p@\hbox{\uwavefont\textcolor{blue}{\char58}}}\ULon}

\newcommand\remove{\bgroup\markoverwith{\textcolor{red}{\rule[0.5ex]{2pt}{0.4pt}}}\ULon}
\makeatother

\SetKwInput{KwData}{Input}
\SetKwInput{KwResult}{Output}

\SetCommentSty{mycommfont}

\lstdefinelanguage{sparql}{
    basicstyle=\tiny,
    keywordstyle=\color{eclipseKeywords},
    commentstyle=\color{blue}, 
    stringstyle=\color{darkgreen},
    keywords={FILTER, NOT, EXISTS, bound, OPTIONAL, BIND, IF, isIRI, AS, INSERT, UPDATE, DELETE, WHERE, SELECT, UNION, rr,rml,ql,rdfs},
    morecomment=[n]{?}{\ },
    morecomment=[l]\#,
    morestring=[b]",
    morestring=[s]{<}{>},
    frame=lines,
}
\lstdefinelanguage{rdf}{
    basicstyle=\scriptsize,
    keywordstyle=\color{eclipseKeywords},
    commentstyle=\color{gray}, 
    stringstyle=\color{darkgreen},
    keywords={rml, ql, rr, rdfs},
    morecomment=[n]{?}{\ },
    morecomment=[l]\#,
    morestring=[b]",
    morestring=[s]{<}{>},
    frame=lines,
}

\lstdefinelanguage{triples}{
    basicstyle=\tiny,
    keywordstyle=\color{eclipseKeywords},
    commentstyle=\color{darkgreen}, 
    keywords={a, rml, ql, rr, rdfs},
    morecomment=[s]{?}{\ },
    frame=lines,
}

\newcommand{\symAttrUniverse}{\mathcal{A}}
\newcommand{\symTermUniverse}{\mathcal{T}}
\newcommand{\symAllStrings}{\mathcal{S}} 
\newcommand{\symAllIRIs}{\mathcal{I}} 
\newcommand{\symAllLiterals}{\mathcal{L}} 
\newcommand{\symAllBNodes}{\mathcal{B}} 
\newcommand{\symRMLGraph}{G}
\newcommand{\symDataObjUni}{\mathcal{D}}
\newcommand{\symDataAccUni}{\mathcal{Q}}

\newcommand{\symAttrSubSet}{A}
\newcommand{\symProjSet}{P}

\newcommand{\bigDataObject}{D}
\newcommand{\symSetDataset}{\mathcal{D}^\texttt{\tiny ds}\!}
\newcommand{\symSetDatasetX}[1]{\mathcal{D}^\texttt{\tiny ds}_{#1}}
\newcommand{\symSetContentA}{\mathcal{D}^\texttt{\tiny c1}}
\newcommand{\symSetContentAX}[1]{\mathcal{D}^\texttt{\tiny c1}_{#1}}
\newcommand{\symSetContentB}{\mathcal{D}^\texttt{\tiny c2}}
\newcommand{\symSetContentBX}[1]{\mathcal{D}^\texttt{\tiny c2}_{#1}}
\newcommand{\symDataAcc}{L}
\newcommand{\symDataAccX}[1]{L_{#1}}

\newcommand{\symAttrQueryMap}{\mathbb{P}}
\newcommand{\symDataSeq}{\bar{O}}

\newcommand{\symTermSeqUniverse}{\bar{\symTermUniverse}}

\newcommand{\symJoinAttrPairs}{\mathbb{J}}

\newcommand{\attr}{a}
\newcommand{\subjAttr}{\attr_\textrm{s}}
\newcommand{\predAttr}{\attr_\textrm{p}}
\newcommand{\objAttr}{\attr_\textrm{o}}
\newcommand{\graphAttr}{\attr_\textrm{g}}

\newcommand{\error}{\epsilon}
\newcommand{\mappingTuple}{t}
\newcommand{\symMappingInst}{I}
\newcommand{\mappingRel}{(\symAttrSubSet, \symMappingInst)}
\newcommand{\mappingRelSpec}[1]{(\symAttrSubSet_{#1}, \symMappingInst_{#1})}
\newcommand{\queryExpr}{q}

\newcommand{\dataObject}{d}
\newcommand{\eval}{\mathit{eval}}
\newcommand{\evalX}[1]{\eval_{#1}}
\newcommand{\cbeval}{\mathit{eval}'\!}
\newcommand{\cbevalX}[1]{\eval_{#1}'}
\newcommand{\sourceType}{type}
\newcommand{\sourceTypeX}[1]{type_{#1}}
\newcommand{\sourceTypeTuple}{(\symSetDataset, \symSetContentA\!, \symSetContentB\!, \symDataAcc, \symDataAcc'\!, \eval, \cbeval, \typeCast)}
\newcommand{\sourceTypeTupleX}[1]{\sourceTypeTupleXXX{#1}{#1}{#1}}
\newcommand{\sourceTypeTupleXXX}[3]{(\symSetDatasetX{#1}, \symSetContentAX{#2}, \symSetContentBX{#3}, \symDataAccX{#2}, \symDataAccX{#3}', \evalX{#2}, \cbevalX{#3}, \typeCastX{#1})}
\newcommand{\dataSource}{s}
\newcommand{\dataSourceTuple}{(\sourceType,\bigDataObject)}

\newcommand{\sourceOp}[3]{\textrm{Source}^{(#1,#2,#3)}}
\newcommand{\sourceOpDflt}{\sourceOp{\dataSource}{\queryExpr}{\symAttrQueryMap}}

\newcommand{\typeCast}{cast}
\newcommand{\typeCastX}[1]{\typeCast_{#1}}

\newcommand{\var}{v}

\newcommand{\extExpr}{\varphi}

\newcommand{\extFunc}{f}
\newcommand{\extFuncType}{(\symTermUniverse \cup \{\error\})}
\newcommand{\extFuncTuple}{(\extFunc,\extExpr_1,\dots,\extExpr_n)}
\newcommand{\extOp}{\text{Extend}_{\varphi}^{\attr}}
\newcommand{\extOpAttr}[1]{\text{Extend}_{\varphi}^{#1}}

\newcommand{\equiJoinOp}{\text{EqJoin}^{\symJoinAttrPairs}}
\newcommand{\triplesMapIRI}{u}
\newcommand{\toIRI}{\texttt{toIRI}}
\newcommand{\toBNode}{\texttt{toBNode}^{\LTB}}
\newcommand{\toLiteral}{\texttt{toLiteral}}

\newcommand{\concat}{\texttt{concat}}

\newcommand{\baseIRI}{\mathit{base}}
\newcommand{\literal}{\ell}
\newcommand{\lex}{\mathit{lex}}
\newcommand{\dt}{\mathit{dt}}
\newcommand{\literalTuple}{(\lex, \dt)}
\newcommand{\LTB}{\mathit{S2B}}
\newcommand{\tempSubStrs}{\textsc{Split}}
\newcommand{\substrSeq}{\bar{S}}
\newcommand{\projectOp}[1]{\text{Project}^{#1}}

\newcommand{\union}{\text{Union}}

\newcommand{\fctDom}[1]{\mathrm{dom}(#1)}

\newcommand{\fctAttrs}[1]{\mathrm{attrs}(#1)}

\newcommand{\fctEval}[2]{\eval(#1,#2)}

\newcommand{\fctCbeval}[3]{\cbeval(#1,#2,#3)}

\newcommand{\fctTypeCast}[1]{\typeCast(#1)}

\newcommand{\fctExtFuncEvalN}[2]{\extFunc\bigl(\fctEval{#1_1}{#2},\dots,\fctEval{#1_n}{#2}\bigr)}
\newcommand{\fctExtOp}[1]{\extOp(#1)}
\newcommand{\fctExtOpBig}[1]{\extOp\bigl(#1\bigr)}
\newcommand{\fctExtOpX}[3]{\text{Extend}^{#1}_{#2}(#3)}
\newcommand{\fctExtOpXBig}[3]{\text{Extend}^{#1}_{#2}\bigl(#3\bigr)}

\newcommand{\fctEquiJoinOp}[2]{\equiJoinOp(#1, #2)}
\newcommand{\fctEquiJoinOpBig}[2]{\equiJoinOp\bigl(#1, #2\bigr)}
\newcommand{\fctToIRI}[2]{\toIRI(#1,#2)} 
\newcommand{\fctToBNode}[1]{\toBNode(#1)}
\newcommand{\fctToLiteral}[1]{\toLiteral(#1)}

\newcommand{\fctConcat}[2]{\concat(#1, #2)}
\newcommand{\fctConcatBig}[2]{\concat\bigl( #1, #2 \bigr)}

\newcommand{\fctTempSubStrs}[1]{\tempSubStrs(#1)}
\newcommand{\fctProjectOpDflt}[1]{\fctProjectOp{\symProjSet}{#1}}
\newcommand{\fctProjectOp}[2]{\projectOp{#1}(#2)}
\newcommand{\fctProjectOpBig}[2]{\projectOp{#1}\bigl(#2\bigr)}

\newcommand{\fctUnion}[2]{\union(#1,#2)}
\newcommand{\fctUnionBig}[2]{\union\bigl(#1,#2\bigr)}

\newcommand{\ttl}[1]{\texttt{\small #1}}  
\newcommand{\ttlSmaller}[1]{\texttt{\scriptsize #1}}  

\hypersetup{bookmarksopen=true, citecolor=blue}
\begin{document}												

\title{\texorpdfstring{An Algebraic Foundation \\ for Knowledge Graph Construction}{An Algebraic Foundation for Knowledge Graph Construction}}
\titlerunning{An Algebraic Foundation for Knowledge Graph Construction}

\ExtendedVersion{
\subtitle{
(Extended Version)%
\thanks{This is an extended version of a research paper accepted for the 22nd Extended Semantic Web Conference (ESWC~2025). The extension consists of
	a few additional equivalences in Section~\ref{ssec:equivalences:ProjectionPushing} and the complete content of Section~\ref{ssec:equivalences:ExtendPushingOrPulling}, which are not included in the version prepared for the conference proceedings.}
}
}

%
\author{%
	Sitt Min Oo\inst{1}\orcidlink{0000-0001-9157-7507}
	\and
	Olaf Hartig\inst{2}\orcidlink{0000-0002-1741-2090}%
}
\authorrunning{S. Min Oo and O. Hartig}
%
\institute{
	Ghent University - imec, Ghent, Belgium \\
	\email{x.sittminoo@ugent.be}
	\and
	Linköping University, Linköping, Sweden \\
	\email{olaf.hartig@liu.se}
}
\maketitle              

\vspace{-5mm} 

\begin{abstract}

Although they exist since more than ten years already, have attracted diverse implementations, and have been used successfully in a significant number of applications, declarative mapping languages for constructing
knowledge graphs from heterogeneous types of data sources still lack a solid formal foundation.
	This
%
	makes it impossible to introduce implementation and optimization techniques that are provably correct and, in fact, has led to discrepancies between different implementations. Moreover,
it precludes
	studying fundamental properties of different languages (e.g., expressive power).
To address this gap, this paper introduces a
	language-agnostic
algebra
	for capturing mapping definitions.
As further contributions, we show that the popular mapping language RML can be translated into our algebra%
	~(%
%
	by which we also provide
%
	a
formal definition of the semantics of RML%
%
	)
and we prove several algebraic
	rewriting rules that can be used
to optimize mapping plans based on our algebra.

\end{abstract}

\vspace{-6mm} 
\enlargethispage{\baselineskip}

\section{Introduction}%
\label{sec:introduction}

\vspace{-2mm} 

Knowledge graphs (KGs) are often created or populated by extracting and transforming data from other sources of structured or semi-structured data. One approach to this end is to
	employ a mapping engine for which the desired mapping from the source data to the KG can be specified using a declarative language.
		This approach has gained traction in recent years, with applications across diverse domains such as COVID research~\cite{Sakor2023KG4Covid}, railway networks~\cite{rojas2021leveraging}, and incident management in ICT systems~\cite{Tailhardat2024NORIA}. To provide a basis for this approach, several declarative mapping languages, in particular for RDF-based KGs, have been proposed,
including languages that focus only on a single type of input data sources (e.g., R2RML for relational databases~\cite{Seema2012R2RML}) and, more interestingly, also languages designed for multiple types of input data sources~\cite{VanAssche2022DeclarativeRDFgraph}. The latter can be classified further into mapping-focused extensions of query languages and dedicated languages
designed solely for the declarative description of mappings.%


So far,
	most
of these mapping languages---in particular, the dedicated ones---are defined only informally. Yet, such an informal definition can easily lead to discrepancies between different
	implementations.
For instance, for
RML~\cite{dimou_ldow_2014,Dimou2024:RMLSpec}, which has
several implementations~\cite{ArenasGuerrero2022MorphKGCScalable,Freund2024FlexRML,Iglesias2022ScalingKnowledgeGraph,MinOo2022RMLStreamer,Scrocca2020Transport}, the introduction of
conformance tests~\cite{Heyvaert2019:RMLTests} has revealed various cases in which
	different implementations fail
such a test~\cite{Heyvaert2022:RMLTestReports}. It may be argued that the developers of these implementations
	interpreted the informal definition of the language differently.
Without a formal definition it is impossible to introduce implementation and optimization techniques that are provably correct%
	, or
to
	properly compare languages or language features in terms of fundamental properties, such as their expressive~power.
%
%
Therefore, the goal of our work in this paper is to provide a
formal foundation for such declarative mapping languages. Especially, instead of focusing on a single specific language, we aim for a more general formalization approach.

Our main contribution is a language-agnostic algebra to capture definitions of mappings from heterogeneous types of data sources to RDF-based KGs. The basis of this mapping algebra is a variation of the relational data model~(Section~\ref{sec:data_model}), for which we define five types of operators that can be arbitrarily combined into algebra expressions~(Section~\ref{sec:mapping_algebra}).
Due to its language-agnostic nature, the algebra can serve as a foundation to
various mapping languages formally. As our second contribution, we demonstrate this benefit by providing an algorithm that translates mappings defined using RML into our algebra~(Section~\ref{sec:expressiveness}). Through this algorithm, we show not only that our algebra is at least as expressive as RML, but we also provide a formal definition of the semantics of RML. To the best of our knowledge, this is the first formal approach to capture the RML semantics.
Another important value of having an algebra that captures declarative mapping definitions is that
	it
may be used as the basis of a systematic and well-defined approach to plan and to optimize
	the execution of
	KG construction processes
in mapping engines%
.
Related to this option, we make our third contribution: We show several algebraic
	equivalences which can be used as rewriting rules
to optimize mapping plans that are based on our algebra~(Section~\ref{sec:equivalences}).

\vspace{-2mm} 
\enlargethispage{\baselineskip}

\section{Related Work}%
\label{sec:related_works}

\vspace{-2mm} 

As mentioned above, declarative mapping languages for KG construction can be classified into languages that are extensions of other types of languages (mainly, query languages), with XSPARQL~\cite{Bischof:2012:XSPARQL}, SPARQL-Generate~\cite{lefrancois_eswc_2017}, and
	Facade-X%
~\cite{Daga2021FacadeXOpinionated} being examples of such ``repurposed mapping languages''~\cite{VanAssche2022DeclarativeRDFgraph}, and ``dedicated mapping languages''~\cite{VanAssche2022DeclarativeRDFgraph} that have been created specifically for the purpose of KG construction, with R2RML~\cite{Seema2012R2RML} and RML~\cite{dimou_ldow_2014} as examples.


Given our focus on formal approaches to capture such languages, for the repurposed ones we notice
they can rely on the formalization of their underlying language.
	For instance,
XSPARQL combines the semantics of XQuery and SPARQL for mapping between XML and RDF data~\cite{Bischof:2012:XSPARQL}, and to also support
	relational databases,
the authors extend their work by utilizing the semantics of relational algebra~\cite{Lopes2011XSPARQL}. Similarly, the formalization of SPARQL-Generate~\cite{lefrancois_eswc_2017} builds on the algebra-based formalization of SPARQL and, thus, is similar in spirit to
	our approach,
which
	we see as more relevant for
dedicated mapping~languages.


When considering dedicated mapping languages, to the best of our knowledge, only R2RML has a formally defined semantics. That is, Sequeda et al.~\cite{Sequeda2013R2RML} provide such a formalization by means of 57 Datalog rules. As this definition is for R2RML, it is limited to mappings from relational databases to RDF, whereas our formalization in this paper can capture mappings of different types of sources of structured and semi-structured data. Moreover, we argue that an algebraic formalization such as ours has the additional advantage that it can also be utilized to represent a form of logical execution plans in mapping engines.

Besides R2RML, a few studies with formalizations related to RML have been published in recent years~\cite{Iglesias2022ScalingKnowledgeGraph,MinOo2023TowardsAlgebra}.
Iglesias et al.\
	formalize
RML-based mapping processes based on Horn clauses~\cite{Iglesias2022ScalingKnowledgeGraph}, where the focus of this formalization is to define optimization techniques presented by the authors.
Lastly, an initial study to capture the semantics of mappings in a language-agnostic manner through algebraic mapping operators~\cite{MinOo2023TowardsAlgebra} has a very similar goal as our work in this paper. Yet, in contrast to our work, the authors do not cover all aspects of
	popular
dedicated mapping language such as RML (e.g., joining data from multiple sources) and several of their definitions remain informal or rely on undefined concepts.

\vspace{-2mm} 
\enlargethispage{\baselineskip}

\section{Data Model}
\label{sec:data_model}

\vspace{-2mm} 

This section introduces the data model
based on which our mapping algebra is defined.
Hence, this model captures the types of intermediate results of any mapping process
described by our formalism.
Informally, the data model is a version of the relational data model, restricted to a special kind of relations, that we call \emph{mapping relations}%
\ and that contain RDF terms as possible values.

As a preliminary step for defining the model formally, we introduce relevant concepts of RDF:
We let $\symAllStrings$ be the countably infinite set of strings (sequences of
Unicode code points) and $\symAllIRIs$ be the subset of $\symAllStrings$ that
consists of all strings that are valid IRIs.
Moreover, $\symAllLiterals$ is a countably infinite set of pairs
$ \literalTuple \in \symAllStrings \times \symAllIRIs$, which we
call \emph{literals}.
For each such literal~$\literal = \literalTuple$, $\lex$ is
its \emph{lexical form} and $\dt$ is its \emph{datatype IRI}.\footnote{To avoid overly complex formulas in this paper, we ignore the option of literals that have a language tag. Yet, the formalism can easily be extended to cover
	them.}
Furthermore, we assume a countably infinite set~$\symAllBNodes$ of \emph{blank
	nodes}, which is disjoint from both $\symAllStrings$ and $\symAllLiterals$.
The set~$\symTermUniverse$ of all \emph{RDF terms} is defined as
$\symTermUniverse = \symAllIRIs \cup \symAllBNodes \cup \symAllLiterals$, and
we write $\symTermSeqUniverse$ to denote the set of all possible sequences of
terms; i.e., each element of $\symTermSeqUniverse$ is a sequence of elements of
$\symTermUniverse$.
As usual, an \emph{RDF triple} is a tuple $(s,p,o) \in (\symAllIRIs \cup \symAllBNodes) \times \symAllIRIs \times \symTermUniverse$, an \emph{RDF graph} is a set of RDF triples, and an \emph{RDF dataset} is a set $\big\{ G_\mathrm{dflt}, (n_1,G_1), \ldots, (n_m,G_m) \big\}$ where $G_\mathrm{dflt}, G_1, \ldots, G_m$ are RDF graphs, $m \geq 0$, and $n_i \in \symAllIRIs \cup \symAllBNodes$ for all $i \in \{1,\ldots,m\}$.

As additional ingredients for defining mapping relations, we assume a countably infinite set~$\symAttrUniverse$ of \emph{attributes} and a special value $\error$
that is not an RDF term~(i.e., $\error \notin \symTermUniverse$) and that shall be used to capture processing errors. The last ingredient
are  \emph{mapping tuples}, which are the type of tuples used in mapping relations.

\begin{definition}
	\normalfont
	A \textbf{mapping tuple} is a partial function $\mappingTuple:
		\symAttrUniverse \rightarrow \symTermUniverse \cup \{\error\}$.
\end{definition}

We say that two
mapping tuples $\mappingTuple$ and $\mappingTuple'$ are \emph{compatible} if, for every attribute $\attr \in \fctDom{\mappingTuple} \cap \fctDom{\mappingTuple'}$, it holds that $\mappingTuple(\attr) = \mappingTuple'(\attr)$.
Given two mapping tuples $\mappingTuple$ and $\mappingTuple'$ that are compatible, we write $\mappingTuple \cup \mappingTuple'$ to denote the mapping tuple $\mappingTuple''$ that is the merge of $\mappingTuple$ and $\mappingTuple'$; that is, $\fctDom{\mappingTuple''} = \fctDom{\mappingTuple} \cup \fctDom{\mappingTuple'}$ and, for every attribute~$\attr \in \fctDom{\mappingTuple''}$,
%
$\mappingTuple''(\attr) = \mappingTuple(\attr)$ if $\attr \in \fctDom{\mappingTuple}$, and $\mappingTuple''(\attr) = \mappingTuple'(\attr)$ otherwise.

Now we are ready to define our notion of a mapping relation.

\begin{definition}
	\label{def:mapping_relation}
	\normalfont
	A \textbf{mapping relation}~$r$ is a tuple $\mappingRel$, where
	$\symAttrSubSet \subset \symAttrUniverse$ is a finite, non-empty set of attributes
	and
	$\symMappingInst$ is a set of mapping tuples such
	that, for every such tuple~$\mappingTuple \in \symMappingInst$, it holds that $\mathrm{dom}(\mappingTuple) = \symAttrSubSet$.
	We call $\symAttrSubSet$ the \textbf{schema} of the mapping relation
	and $\symMappingInst$ is the \textbf{instance} of the mapping relation.
\end{definition}

For the sake of conciseness, we sometimes write $\mappingTuple \in r$ to denote that $\mappingTuple$ is a mapping tuple in
the instance~%
$\symMappingInst$ of a mapping relation $r=\mappingRel$.

\begin{table}[t]
	\centering
	\caption{Mapping relation~%
		$r$ discussed in Examples~\ref{ex:mapping_relation} and~\ref{ex:generated_dataset}.}
\vspace{-2mm} 
	\label{tab:mapping_relation}
	\begin{tabular}{c|c|c|c|c|c|}
		\cline{2-6}
		                  & $\attr_\mathrm{x}$                & $\subjAttr$    & $\predAttr$      & $\objAttr$                        & $\graphAttr$
		\\ \cline{2-6}
		$\mappingTuple$   & $(\ttlSmaller{"12"}, \ttlSmaller{xsd:integer})$ & \ttlSmaller{ex:alice} & \ttlSmaller{foaf:knows} & \ttlSmaller{ex:bob}                      & \ttlSmaller{rr:defaultGraph}
		\\
		$\mappingTuple'$  & \ttlSmaller{ex:alice}                    & \ttlSmaller{ex:alice}       & $(\ttlSmaller{"knows"}, \ttlSmaller{xsd:string})$ & \ttlSmaller{ex:charles} & \ttlSmaller{ex:g1}
		\\
		$\mappingTuple''$ & $\error$                          & \ttlSmaller{ex:bob}   & \ttlSmaller{foaf:name}  & $(\ttlSmaller{"Bob"}, \ttlSmaller{xsd:string})$ & \ttlSmaller{ex:g2}
		\\ \cline{2-6}
	\end{tabular}
\vspace{-2mm} 
\end{table}
\begin{example}
	\label{ex:mapping_relation}
	Consider the mapping relation~$r = \mappingRel$ that is represented in a tabular form in Table~\ref{tab:mapping_relation}.
	It holds that $\symAttrSubSet = \{ \attr_\mathrm{x}, \subjAttr, \predAttr, \objAttr, \graphAttr \}$ and
	$\symMappingInst$ contains three mapping tuples, $\mappingTuple$, $\mappingTuple'\!$, and $\mappingTuple''\!$, such that
	$\mappingTuple(\attr_\mathrm{x})$ is the literal $(\ttl{"12"},\ttl{xsd:integer})$,
	$\mappingTuple'(\attr_\mathrm{x})$ is the IRI \ttl{ex:alice},
	$\mappingTuple''(\attr_\mathrm{x}) = \error$, etc.
\end{example}

While mapping relations are
intermediate results of mapping processes described by our algebra, the ultimate goal of
such processes
is to produce an RDF dataset. Therefore, we also need to define how mapping relations are mapped into RDF datasets. To this end, we assume four special attributes: $\subjAttr,\predAttr,\objAttr,\graphAttr \in \symAttrUniverse$. Every mapping relation whose schema contains these attributes is mapped to an RDF dataset
by using the values that the tuples in the relation have for these attributes. That is, if the values that a mapping tuple has for $\subjAttr$, $\predAttr$, and $\objAttr$ form a valid RDF triple, then the graph determined by the $\graphAttr$-value of the tuple contains this triple. Formally, we define this approach as follows.

\begin{definition}
	\label{def:generated_dataset}
	\normalfont
	Let $r = \mappingRel$ be a mapping relation with $\subjAttr, \predAttr, \objAttr, \graphAttr \in \symAttrSubSet$, and
	\begin{align*}
		\symMappingInst_\mathrm{valid}
		 & =
		\bigl\{
		\mappingTuple \in \symMappingInst
		\,\big|\,
		\bigl( \mappingTuple(\subjAttr), \mappingTuple(\predAttr), \mappingTuple(\objAttr) \bigr) \text{ is an RDF triple and } \mappingTuple(\graphAttr) \in \symAllIRIs\cup\symAllBNodes
		\bigr\}
		\; \text{ and}
		\\
		N
		 & =
		\bigl\{ \mappingTuple(\graphAttr) \in \symAllIRIs\cup\symAllBNodes \,\big|\, \mappingTuple \in \symMappingInst_\mathrm{valid} \text{ and $\mappingTuple(\graphAttr)$ is not the IRI } \ttl{rr:defaultGraph} \bigr\}
		.
	\end{align*}
	The \textbf{RDF dataset resulting from $r$} is the RDF dataset
	\begin{equation*}
		D =
		\{ G_\mathrm{dflt} \}
		\cup
		\bigl\{
		( n, G_n )
		\,\big|\,
		n \in N
		\bigr\}
		,
	\end{equation*}
	where, for every $n \in N$,
	\begin{align*}
		G_n             & =
		\bigl\{
		\bigl( \mappingTuple(\subjAttr), \mappingTuple(\predAttr), \mappingTuple(\objAttr) \bigr)
		\,\big|\,
		\mappingTuple \in \symMappingInst_\mathrm{valid} \text{ such that } \mappingTuple(\graphAttr)=n
		\bigr\}
		\; \text{ and}
		\\
		G_\mathrm{dflt} & =
		\bigl\{
		\bigl( \mappingTuple(\subjAttr), \mappingTuple(\predAttr), \mappingTuple(\objAttr) \bigr)
		\,\big|\,
		\mappingTuple \in \symMappingInst_\mathrm{valid} \text{ such that } \mappingTuple(\graphAttr) \text{ is } \ttl{rr:defaultGraph}
		\bigr\}
		.
	\end{align*}
\end{definition}

\begin{example}
	\label{ex:generated_dataset}
	For the mapping relation in Example~\ref{ex:mapping_relation} it holds that $\symMappingInst_\mathrm{valid} = \{ \mappingTuple, \mappingTuple'' \}$ and $N = \{ \ttl{ex:g2} \}$. Therefore, the RDF dataset resulting from this mapping relation
	is $D = \bigl\{ G, (\ttl{ex:g2}, G') \bigr\}$
	with $G = \big\{ (\ttl{ex:alice}, \ttl{foaf:knows}, \ttl{ex:bob}) \big\}$
	and $G'\! = \big\{ (\ttl{ex:bob}, \ttl{foaf:name}, (\ttl{"Bob"},\ttl{xsd:string})) \big\}$. Notice that $\mappingTuple'\! \notin \symMappingInst_\mathrm{valid}$ because $\big( \mappingTuple'(\subjAttr), \mappingTuple'(\predAttr), \mappingTuple'(\objAttr) \bigr)$ is not an RDF triple%
		~(since $\mappingTuple'(\predAttr)$ is a literal)%
	.
\end{example}

\vspace{-2mm} 
\enlargethispage{\baselineskip}

\section{Mapping Algebra}%
\label{sec:mapping_algebra}



This section introduces our mapping algebra which consists of
	five
types of operators.
The output of each such operator is a mapping relation and the inputs (if any)
are mapping relations as well.
Hence, the operators can be combined into arbitrarily complex algebra
expressions, where each such expression defines a mapping from one or more
sources of (semi-)structured data into an~RDF~dataset.\footnote{We assume here
that the mapping relation resulting from the root operator of the expression
contains the aforementioned attributes $\subjAttr$, $\predAttr$, $\objAttr$,
and $\graphAttr$ (see Definition~\ref{def:generated_dataset}).}

\subsection{Source Operator}%
\label{sub:source_operator}

\vspace{-2mm} 
\enlargethispage{\baselineskip}

The basis of every expression that defines a mapping using our algebra are
operators that establish a relational view of each source of input data. 
From the perspective of the algebra, these \emph{source operators} are nullary
operators; they do not have any mapping relation as input. 
The mapping relation that such an operator creates as output depends on the
data of the corresponding data source and can be specified based on a language
appropriate to identify and extract relevant pieces of data from the data
source. 
As our mapping algebra should be applicable for various types of data sources,
for which different such languages are suitable, we introduce an abstraction of
this extraction process and define our notion of source operators based on this
abstraction.

Our abstraction considers
two infinite sets, $\symDataObjUni$ and $\symDataAccUni$, which are disjoint.
$\symDataObjUni$ is the set of all possible \emph{data objects}.
In practice,
there are different kinds of data objects, even within the context of the same type of data source, and data objects may be nested within one another. For instance,
each line of a CSV file may be seen as a data object, and so may each value within such a line, as well as the CSV file as a whole.
In our abstraction, any particular kind of data objects is considered as a specific subset of $\symDataObjUni$%
.

The set $\symDataAccUni$ captures possible \emph{query languages} to retrieve data objects from other data objects. We consider each such language $\symDataAcc \in \symDataAccUni$ as a set, where every element $\queryExpr \in \symDataAcc$ is one of the expressions
admitted by
the language. In practice, each such language is designed to be used for a specific kind of data objects (i.e., a specific subset of $\symDataObjUni$).
Therefore, we define the notion of the \emph{types of data sources} considered for the source operators of our algebra as a structure that specifies the relevant kinds of data objects and the corresponding evaluation semantics of the languages to be used for data sources of such a type.

\vspace{-1mm} 

\begin{definition}
	\label{def:source_type}
	\normalfont
	A \textbf{source type} is a tuple $\sourceTypeTuple$
	where%
\vspace{-2mm} 
	\begin{itemize}
		\item
		      $\symSetDataset \subseteq \symDataObjUni$,
		      \, $\symSetContentA\! \subseteq \symDataObjUni$,
		      \, $\symSetContentB\! \subseteq \symDataObjUni$,

		\item
		      $\symDataAcc$ and $\symDataAcc'$ are query languages in $\symDataAccUni$,

		\item
		      $\eval$ is a function $\eval \!: \symSetDataset \times \symDataAcc \rightarrow \overline{\symSetContentA}$,

		\item
		      $\cbeval$ is a partial function $\cbeval : \symSetDataset \times \symSetContentA\! \times \symDataAcc'\! \rightarrow \overline{\symSetContentB}$, and

		\item
		      $\typeCast$ is a partial function $\typeCast \!: \symSetContentB\! \rightarrow \symAllLiterals$,
	\end{itemize}
\vspace{-2mm} 
	where $\overline{\symSetContentA}$, resp.\ $\overline{\symSetContentB}$, is the set of all
	sequences of data objects in $\symSetContentA\!$, resp.\ $\symSetContentB\!$.
\end{definition}

\vspace{-4mm} 

\begin{definition}
	\label{def:data_source}
	\normalfont
	A \textbf{data source}~$\dataSource$ is a tuple~$\dataSourceTuple$ in which
	$\sourceType$ is a source type
	$(\symSetDataset, \mathcal{D}^\texttt{\tiny o1}\!, \mathcal{D}^\texttt{\tiny o2}\!, \symDataAcc, \symDataAcc'\!, \eval, \cbeval, \mathit{cast})$
	and
	$\bigDataObject \in \symSetDataset$.
\end{definition}

\vspace{-1mm} 

As per Definitions~\ref{def:source_type}%
\ and~%
\ref{def:data_source}, the set $\symSetDataset$ of a source type
determines the kind of data objects that the data sources of this type may provide access to.
For instance, for CSV-based data sources this may be the (infinite) set of all possible CSV files.
The
functions~$\eval$ and $\cbeval$ are meant to define the evaluation semantics of the languages~$\symDataAcc$ and~$\symDataAcc'\!$, respectively.
Notice that the domain of $\cbeval$ consists of 3-tuples that contain an additional data object as their second element, which is meant as a context object for the evaluation. The idea of the source operators shall be to use $\symDataAcc$ to specify relevant context objects and, then, to use $\symDataAcc'$ to retrieve values for
creating a mapping tuple
%
per
context~object.

While a formal definition of concrete source types is beyond the scope of this paper, the
following two examples outline informally how JSON-based data sources and CSV-based data sources may be captured within our abstraction.

\vspace{-1mm} 

\begin{example}
	\label{ex:source_type_json}
	To define $\sourceTypeX{\mathsf{json}} = \sourceTypeTupleXXX{\mathsf{json}}{\mathsf{jp}}{\mathsf{jp}}$ as a source type for JSON-based data sources, we let $\symSetDatasetX{\mathsf{json}}$ be
	the infinite set of all possible JSON documents, both $\symSetContentAX{\mathsf{jp}}$ and $\symSetContentBX{\mathsf{jp}}$ are the infinite set of all possible elements that may occur within JSON documents (i.e., $\symSetContentAX{\mathsf{jp}} = \symSetContentBX{\mathsf{jp}}$),
	$\symDataAccX{\mathsf{jp}}$ is the set of all JSONPath expressions that begin with the selector for the root element (i.e., \texttt{\$}), and
	$\symDataAccX{\mathsf{jp}}'$ are all JSONPath expressions that do not begin with~\texttt{\$}.
	Then, $\evalX{\mathsf{jp}}$ and $\cbevalX{\mathsf{jp}}$ capture the evaluation semantics of JSONPath.
	That is, for a JSON document $\bigDataObject \in \symSetDatasetX{\mathsf{json}}$ and a query $\queryExpr \in \symDataAccX{\mathsf{jp}}$, $\evalX{\mathsf{jp}}(\bigDataObject,\queryExpr)$ returns the elements of $\bigDataObject$ selected by $\queryExpr$. If $\queryExpr'\! \in \symDataAccX{\mathsf{jp}}'$, and assume $\dataObject \in \symSetContentAX{\mathsf{jp}}$ is a JSON object within $\bigDataObject$, then $\cbevalX{\mathsf{jp}}(\bigDataObject,\dataObject,\queryExpr')$ returns the elements of $\bigDataObject$ that can be reached from $\dataObject$ as per the traversal specified by $\queryExpr'\!$.
	Finally, $\typeCastX{\mathsf{json}}$ maps JSON values to RDF literals. For instance, string values may be mapped to \ttl{xsd:string} literals, numeric values that are integers may be mapped to \ttl{xsd:integer} literals, etc.
\end{example}

\vspace{-3mm} 

\begin{example}
	\label{ex:source_type_csv}
	As a
	source
	type for CSV-based data sources, we may
	introduce
	$\sourceTypeX{\mathsf{csv}} = \sourceTypeTupleX{\mathsf{csv}}$ as follows.
	$\symSetDatasetX{\mathsf{csv}}$ is the infinite set of all possible CSV files,
	$\symSetContentAX{\mathsf{csv}}$ is the infinite set of all possible rows of CSV files,
	$\symSetContentBX{\mathsf{csv}} \subset \symAllStrings$ is the infinite set of strings that may be values within CSV files, and
	the $\typeCastX{\mathsf{csv}}$ function maps all these string values
	in $\symSetContentBX{\mathsf{csv}}$
	to \ttl{xsd:string} literals.
	As the language $\symDataAccX{\mathsf{csv}}$, we may use the singleton set that contains (only) the empty string~$\varepsilon$, and $\symDataAcc'_{\mathsf{csv}} \subset \symAllStrings$ is the set of all strings that may be used as column names in CSV files. Informally, $\evalX{\mathsf{csv}}$ and $\cbevalX{\mathsf{csv}}$ may then be defined as follows. For
	every CSV file~$\bigDataObject \in \symSetDatasetX{\mathsf{csv}}$, $\evalX{\mathsf{csv}}(\bigDataObject,\varepsilon)$ returns all rows of~$\bigDataObject$%
	, and for every such row
	$\dataObject \in \evalX{\mathsf{csv}}(\bigDataObject,\varepsilon)$
	and every string $c$ that is a column name in~$\bigDataObject$, $\cbevalX{\mathsf{csv}}(\bigDataObject,\dataObject,c)$ is the value that
	the row~%
	$\dataObject$ has in column~$c$.
\end{example}

\vspace{-3mm} 

\begin{example}
	\label{ex:running_ex_source_type}
	As the basis of a running example for the rest of this section, we assume a
	data source $\dataSource_\mathsf{ex} = (\sourceTypeX{\mathsf{csv}},\bigDataObject_\mathsf{ex})$ where $\bigDataObject_\mathsf{ex}$ is a CSV file with four columns, named \textsf{\small id}, \textsf{\small firstname},
	\textsf{\small lastname}, and \textsf{\small age}, and the following two rows:%
\vspace{-2mm} 
	\begin{equation*}
        (\ttl{1}, \ttl{Alice}, \ttl{Lee}, \ttl{23}) \quad \text{ and } \quad (\ttl{2}, \ttl{Bob}, \ttl{Malice}, \ttl{unknown})
        .
\vspace{-2mm} 
	\end{equation*}
	If we write $\dataObject_1$ to denote the first of these rows,
		then
	$\cbevalX{\mathsf{csv}}(\bigDataObject_\mathsf{ex}, \dataObject_1, \textsf{\small firstname})$ is the string
	\ttl{Alice}, and $\typeCastX{\mathsf{csv}}(\ttl{Alice})$ is the RDF literal $(\ttl{"Alice"}, \ttl{xsd:string})$. 
\end{example}

\vspace{-1mm} 
\enlargethispage{\baselineskip}

Now we are ready to define our notion of source operators. While they are nullary operators, they are specified using three parameters:
the data source from which the data for the produced mapping relation is extracted,
a query that selects relevant context objects from the data source,
and a map that associates attributes with
queries, where these queries are meant to select the values that a mapping tuple produced for a context object has for the attributes.
Formally, we define the mapping relation produced by a source operator as follows.

\vspace{-1mm} 

\begin{definition}
	\label{def:source_operator}
	\normalfont
	Let $\dataSource = \dataSourceTuple$ be a data source with
		$\sourceType = (\symSetDataset, \symSetContentA\!, \symSetContentB\!,$ $\symDataAcc, \symDataAcc'\!, \eval, \cbeval, \typeCast)$,
		let
	$\queryExpr \in \symDataAcc$, and
		let
	$\symAttrQueryMap$ be a partial function
	$\symAttrQueryMap \!: \symAttrUniverse  \rightarrow \symDataAcc'\!$.
	The \textbf{$(\queryExpr, \symAttrQueryMap)$-spe\-cif\-ic mapping relation obtained from $\dataSource$}, denoted by $\sourceOpDflt$, is the mapping relation $\mappingRel$ such that $\symAttrSubSet = \fctDom{\symAttrQueryMap}$ and

	\begin{align*}
		\symMappingInst = \big\{ \{\attr_1 \rightarrow \fctTypeCast{\var_1}, \dots, \attr_n \rightarrow \fctTypeCast{\var_n} \} \ \big| \
		 & \dataObject \text{ is in } \symDataSeq \text{ and}  
		\\[-1mm]
		 & ((\attr_1 , \var_1), \dots, (\attr_n , \var_n)) \in X_d \big\}
		,
\\[-6mm] 
	\end{align*}
	where $\symDataSeq = \fctEval{\bigDataObject}{\queryExpr}$ and%
\vspace{-2mm} 
	\begin{equation*}
		X_d = \bigtimes_{\attr \in \fctDom{\symAttrQueryMap}}
		\big\{ (\attr, \var) \ \big| \
		\var \text{ is in } \fctCbeval{\bigDataObject}{\dataObject}{\queryExpr'} 
		\text{ with }  \queryExpr'\! = \symAttrQueryMap(\attr) \big\}
		.
	\end{equation*}
\end{definition}

\begin{example}
	\label{ex:source_operator}
	The result of $\sourceOp{\dataSource_\mathsf{ex}}{\varepsilon}{\symAttrQueryMap_\mathsf{ex}}$ with
		source
	$\dataSource_\mathsf{ex}$ of Example~\ref{ex:running_ex_source_type} and
		$\symAttrQueryMap_\mathsf{ex} =$ $\{\attr_1 \rightarrow \textsf{\small id}, \attr_2 \rightarrow \textsf{\small firstname}, \attr_3 \rightarrow \textsf{\small age}\}$
	is the
	relation $\big(\{ \attr_1, \attr_2, \attr_3\}, \{\mappingTuple_1,\mappingTuple_2\}\big)$~with%
\vspace{-2mm} 
	\begin{align*}
		\mappingTuple_1 = \big\{
		& \attr_1 \rightarrow (\ttl{"1"}, \ttl{xsd:string}),
		\quad \text{~~~~and}
		&
		\mappingTuple_2 = \big\{
		& \attr_1 \rightarrow (\ttl{"2"}, \ttl{xsd:string}),
		\\[-1mm]
		& \attr_2 \rightarrow (\ttl{"Alice"}, \ttl{xsd:string}),
		&
		& \attr_2 \rightarrow (\ttl{"Bob"}, \ttl{xsd:string}),
		\\[-1.5mm]
		& \attr_3 \rightarrow  (\ttl{"23"}, \ttl{xsd:string}) \big\}
		&
		& \attr_3 \rightarrow (\ttl{"unknown"}, \ttl{xsd:string}) \big\}
		.
	\end{align*}
\end{example}

\vspace{-4mm} 
\enlargethispage{\baselineskip}

\subsection{Extend Operator}%
\label{sub:extend_operator}

\vspace{-2mm} 

Extend operators are unary operators that
	can be used to
extend a mapping relation with an additional attribute. The values that the tuples of the relation shall have for this attribute are specified
	via
a so-called \emph{extend expression} that can be formed using \emph{extension functions}. In the following, we first introduce the notions of such functions and such expressions, and then define extend~operators.

\begin{definition}
	\label{def:ExtensionFunction}
	\normalfont
		An \textbf{extension function} is a function $\extFunc$ of the form
\vspace{-2mm} 
		$$\extFunc \!: \extFuncType \times \dots \times \extFuncType \rightarrow \extFuncType.$$
\end{definition}

\begin{example}
	\label{ex:ExtensionFunction:toInt}
	A concrete example of such an extension function may be a unary function called \texttt{toInt} that converts \ttl{xsd:string} literals representing integers into \ttl{xsd:integer} literals. Formally, for every $v \in (\symTermUniverse \cup \error)$, we define:%
\vspace{-2mm} 
	\begin{equation*}
		\texttt{toInt}(v) \!=\!
		\begin{cases}
			(\lex, \ttl{xsd:integer})
			& \text{if $v$ is a literal $\literalTuple$ such that $\lex$ is in the}
			\\[-1mm]
			& \text{lexical space of the datatype denoted by the}
			\\[-1mm]
			& \text{IRI \ttl{xsd:integer} and $\dt$ is the IRI \ttl{xsd:string},}
			\\
			\error
			& \text{else.}
		\end{cases}
	\end{equation*}
\end{example}


%

\begin{definition}
	\label{def:ExtendExpressions:Syntax}
	\normalfont
		\textbf{Extend expressions} are defined recursively as follows:%
\vspace{-2mm} 
	\begin{enumerate}
		\item
		Every RDF term in $\symTermUniverse$ is an extend expression.

		\item
		Every attribute in $\symAttrUniverse$ is an extend expression.

		\item
		If $\extExpr_1,\dots,\extExpr_n$ are extend expressions and $\extFunc$ is an extension function%
			~(as per Definition~\ref{def:ExtensionFunction})%
		, then the tuple $\extFuncTuple$ is an extend expression.
	\end{enumerate}
\end{definition}

For every extend expression~$\extExpr$, we write $\fctAttrs{\extExpr}$ to denote the set of all attributes mentioned in $\extExpr$. Formally, this set is defined recursively as follows.%
\vspace{-2mm} 
\begin{enumerate}
	\item If $\extExpr$ is an RDF term, then $\fctAttrs{\extExpr} = \emptyset$.
	\item If $\extExpr$ is an attribute $\attr \in \symAttrUniverse$, then $\fctAttrs{\extExpr} = \{\attr\}$.
	\item If $\extExpr$ is of the form $\extFuncTuple$, then $\fctAttrs{\extExpr} = \bigcup_{i \in \{1,\ldots,n\}} \fctAttrs{\extExpr_i}$.
\end{enumerate}

While Definition~\ref{def:ExtendExpressions:Syntax} introduces the syntax of extend expressions, their interpretation as a function to obtain values for mappings tuples is defined as follows.

\begin{definition}
	\label{def:ExtendExpressions:Semantics}
	\normalfont
	Let $\extExpr$ be an extend expression and $\mappingTuple$ be a mapping tuple.
	The \textbf{evaluation of $\extExpr$ over $\mappingTuple$}, denoted by $\fctEval{\extExpr}{\mappingTuple}$, is either an RDF term or the error value ($\error$), and is determined recursively as follows:%
%
%
\vspace{-2mm} 
	\begin{equation*}
		\fctEval{\extExpr}{\mappingTuple} =
		\begin{cases}
			\extExpr
			& \text{if $\extExpr$ is an RDF term,}
			\\
			\mappingTuple(\extExpr)
			& \text{if $\extExpr \in \symAttrUniverse$ and $\extExpr \in \fctDom{\mappingTuple}$,}
			\\
			\fctExtFuncEvalN{\extExpr}{\mappingTuple}
			& \text{if $\extExpr$ is of the form $\extFuncTuple$} \\[-1mm]
			& \text{such that $\extFunc$ is an $n$-ary function,} \\
			\error
			& \text{else.}
		\end{cases}
	\end{equation*}
\end{definition}

\begin{example}
	\label{ex:ExtendExpressions:Semantics}
	If $\extExpr_\mathsf{ex}$ is the extend expression $(\texttt{toInt}, \attr_3)$, for the
    tuples~$\mappingTuple_1$ and~$\mappingTuple_2$ of
    Example~\ref{ex:source_operator}, $\fctEval{\extExpr_\mathsf{ex}}{\mappingTuple_1}$ is
    the literal $(\ttl{"23"}, \ttl{xsd:integer})$ and
    $\fctEval{\extExpr_\mathsf{ex}}{\mappingTuple_2} = \error$.
\end{example}

Now we can define the semantics of extend operators as follows.

\begin{definition}
	\label{def:extend_operator}
	\normalfont
	Let $r = \mappingRel$ be a mapping relation,
		$\attr$ be an attribute that is not in $\symAttrSubSet$ (i.e., $\attr \in \symAttrUniverse \setminus \symAttrSubSet$),
	and $\extExpr$ be an extend expression.
	The \textbf{$(\attr, \extExpr)$-extension of~$r$},
	denoted by $\fctExtOp{r}$,
	is the mapping relation
	$(\symAttrSubSet'\!,\symMappingInst')$ such that
	$\symAttrSubSet'\! = \symAttrSubSet \cup \{ \attr \}$ and
	$\symMappingInst'\! = \bigl\{\mappingTuple \cup \{\attr \rightarrow \fctEval{\extExpr}{\mappingTuple}\} \ | \ \mappingTuple \in \symMappingInst \bigr\}$.
\end{definition}

\begin{example}
	\label{ex:extend_operator}
		We may extend the mapping relation~$r$ of Example~\ref{ex:source_operator}
	with an attribute $\attr_4$ for which each of the
		two tuples of $r$
	gets the value obtained by applying the extend expression~%
		$\extExpr_\mathsf{ex}$
	of Example~\ref{ex:ExtendExpressions:Semantics}. This operation can be captured as $\fctExtOpX{\attr_4}{\extExpr_\mathsf{ex}}{r}$ and results in the mapping relation $\big(\{ \attr_1, \attr_2, \attr_3, \attr_4\}, \{\mappingTuple_1',\mappingTuple_2'\}\big)$ with%
\vspace{-2mm} 
	\begin{equation*}
		\mappingTuple_1' = \mappingTuple_1 \cup \{
        \attr_4 \rightarrow (\ttl{"23"}, \ttl{xsd:integer})
		\}
		\quad \text{ and } \quad
		\mappingTuple_2' = \mappingTuple_2 \cup \{
        \attr_4 \rightarrow \error
		\}
		.
	\end{equation*}
\end{example}

While the example shows how extend operators can be used to convert literals from one datatype to another, other use cases include: adding the same constant RDF term to all tuples of a mapping relation%
, creating IRIs from other attribute values, and applying arbitrary functions to one or more other attribute values.

\vspace{-2mm} 
\enlargethispage{\baselineskip}

\subsection{Relational Algebra Operators}

\vspace{-2mm} 

In addition to the operators introduced above, any operator of the relational algebra can also be used for mapping relations. For instance, the following definition introduces the notion of \emph{projection}, adapted to mapping relations.

\begin{definition}
	\label{def:projection_operator}
	\normalfont
	Let $r = \mappingRel$ be a mapping relation and $\symProjSet \subseteq
		\symAttrSubSet$ be a non-empty subset of the
	attributes of $r$.
	The \textbf{$\symProjSet$-specific projection of $r$}, denoted by $\fctProjectOpDflt{r}$, is the mapping relation
		$(\symProjSet,\symMappingInst')$ such that $\symMappingInst' = \{ \mappingTuple[\symProjSet] \ | \ \mappingTuple \in \symMappingInst \}$,
	where, for every mapping tuple~$\mappingTuple \in \symMappingInst$, $\mappingTuple[\symProjSet]$ denotes the mapping tuple $\mappingTuple'$ that is the restriction of $\mappingTuple$ to $\symProjSet$;
	i.e., $\mathrm{dom}(\mappingTuple') = \symProjSet$ and $\mappingTuple'(a) = \mappingTuple(a)$ for all $a \in \symProjSet$.
\end{definition}

\begin{example}
	\label{ex:projection_operator}
		Let $r$ be the mapping relation of Example~\ref{ex:extend_operator}. Assuming that further steps of a potential mapping process that involve this relation use only the attributes $\attr_2$ and $\attr_4$ of $r$, we may project away the rest of the attributes. To this end, we use $\fctProjectOp{\{\attr_2, \attr_4\}}{r}$ and obtain
	the
	relation $\big(\{ \attr_2, \attr_4\}, \{\mappingTuple_1'',\mappingTuple_2''\}\big)$~with 
\vspace{-2mm} 
	\begin{align*}
        \mappingTuple_1'' &= \big\{ \attr_2 \rightarrow (\ttl{"Alice"}, \ttl{xsd:string}), \, \attr_4 \rightarrow (\ttl{"23"}, \ttl{xsd:integer}) \big\}
        \text{ and}
        \\
        \mappingTuple_2'' &= \big\{ \attr_2 \rightarrow (\ttl{"Bob"}, \ttl{xsd:string}), \, \attr_4 \rightarrow \error \big\}
        .
	\end{align*}
\end{example}

%
%


As two more examples of adopting traditional relational algebra operators for mapping relations, we present the following two definitions.

\begin{definition}
	\label{def:equijoin_operator}
	\normalfont
	Let $r_{1} = \mappingRelSpec{1}$ and
	$r_{2} = \mappingRelSpec{2}$ be mapping relations
	such that $\symAttrSubSet_{1} \cap \symAttrSubSet_{2} = \emptyset$,
	and
	$\symJoinAttrPairs \subseteq \symAttrSubSet_{1} \times \symAttrSubSet_{2}$.
	The \textbf{$\symJoinAttrPairs$-%
		based
	equijoin of $r_{1}$ and $r_{2}$},
	denoted by $\fctEquiJoinOp{r_{1}}{r_{2}}$, is the mapping relation
	$(\symAttrSubSet, \symMappingInst)$
	such that $\symAttrSubSet = \symAttrSubSet_{1} \cup \symAttrSubSet_{2}$
	and%
\vspace{-2mm} 
	\begin{equation*}
		\symMappingInst =  \{ \mappingTuple_1 \cup \mappingTuple_2 \ | \
		\mappingTuple_1 \in \symMappingInst_{1} \text { and }
		\mappingTuple_2 \in \symMappingInst_{2} \text{ such that }
		\mappingTuple_1(\attr_1) = \mappingTuple_2(\attr_2) \text{ for all }
		(\attr_1,\attr_2) \in \symJoinAttrPairs
		\}
		.
	\end{equation*}
\end{definition}

\begin{definition}
	\label{def:union_operator}
	\normalfont
	The \textbf{union} of mapping relations $r_{1} = \mappingRelSpec{1}$ and $r_{2} = \mappingRelSpec{2}$ with $\symAttrSubSet_1 = \symAttrSubSet_2$, denoted by $\fctUnion{r_{1}}{r_{2}}$, is the mapping relation $(\symAttrSubSet_1, \symMappingInst_1 \cup \symMappingInst_2)$.
\end{definition}

\vspace{-4mm} 
\enlargethispage{\baselineskip}

\section{Algebra-Based Definition of RML}%
\label{sec:expressiveness}

\vspace{-2mm} 

This section
	introduces
an algorithm to convert mappings defined using the mapping language RML (specifically, version~1.1.2~\cite{Dimou2024:RMLSpec}) into our
	mapping
algebra.
	This algorithm shows
that \emph{our algebra is at least as expressive as RML}. Moreover, through this algorithm, in combination with the algebra, we provide
	a \emph{%
formal definition of the semantics of RML mappings}.
We emphasize that this formal semantics coincides with the informally-defined semantics of RML~v1.1.2~\cite{Dimou2024:RMLSpec}, which we have verified by running the official RML test cases~\cite{Heyvaert2019:RMLTests} on a prototypical implementation of our approach.\footnote{\url{https://github.com/s-minoo/eswc-2025-poster-algebra-implementation}}
Hereafter, we describe the algorithm step by step, while a complete pseudocode representation is given as Algorithm~\ref{alg:rml2algebra}.
For the description, we assume
	familiarity
with the concepts~of~RML%
	~\cite{dimou_ldow_2014,Dimou2024:RMLSpec,IglesiasMolina2023:RMLOntology}%
.

\begin{algorithm}[t!]
    \SetArgSty{textnormal}
    \DontPrintSemicolon
	\caption{Translates an RML mapping into our mapping algebra.}
	\label{alg:rml2algebra}
        \KwData{%
            $\symRMLGraph$ - an RDF graph (assumed to describe an RML mapping) \;
            \hspace{10mm}
            $\baseIRI$ - an IRI considered as base IRI \;
            \hspace{10mm}
            $\LTB$ - an injective function
            	that maps
            every string to a unique blank~node
        }
        \KwResult{a mapping relation with attributes $\subjAttr$, $\predAttr$, $\objAttr$, and $\graphAttr$}

		$\symRMLGraph_\mathsf{NF} \gets$ normalized version of $\symRMLGraph$ (apply the update queries of Appendix~\ref{appdx:queries} to $\symRMLGraph$) \;

		$\omega_\mathsf{spog} \gets$
			the empty mapping relation $( \{ \subjAttr, \predAttr, \objAttr, \graphAttr\}, \emptyset )$ \;

		\ForEach{$\triplesMapIRI_\mathsf{tm}\! \in \symAllIRIs \cup \symAllBNodes$ for which there exists $o \in \symAllIRIs \cup \symAllBNodes$ such that $\symRMLGraph_\mathsf{NF}$ contains \hspace*{55mm}
		the triple $(\triplesMapIRI_\mathsf{tm},\ttlSmaller{rr:predicateObjectMap},o)$%
		\label{line:rml2algebra:main_loop_begin}
		}
        {
			$\omega \gets \sourceOp{\dataSource}{\queryExpr}{\symAttrQueryMap}$\!,
				where
			$(\dataSource, \queryExpr) = \textsc{SrcAndRootQuery}(\triplesMapIRI_\mathsf{tm},\symRMLGraph_\mathsf{NF})$
			\tcp*[f]{Definition~\ref{def:initSrcAndRootQuery}}
			\hspace*{23mm} and $\symAttrQueryMap = \textsc{ExtractQueries}(\triplesMapIRI_\mathsf{tm},\symRMLGraph_\mathsf{NF})$
			\tcp*{Algorithm~\ref{alg:query_extraction}}
			\label{line:rml2algebra:source}

			$u_\textsf{sm} \gets o$, where $o \in \symAllIRIs \cup \symAllBNodes$ such that $(\triplesMapIRI_\mathsf{tm},\ttlSmaller{rr:subjectMap},o) \in \symRMLGraph_\mathsf{NF}$ \;

			$u_\textsf{pom} \gets o$, where $o \in \symAllIRIs \cup \symAllBNodes$ such that $(\triplesMapIRI_\mathsf{tm},\ttlSmaller{rr:predicateObjectMap},o) \in \symRMLGraph_\mathsf{NF}$ \; 

			$u_\textsf{pm} \gets o$, where $o \in \symAllIRIs \cup \symAllBNodes$ such that $(u_\textsf{pom},\ttlSmaller{rr:predicateMap},o) \in \symRMLGraph_\mathsf{NF}$ \;

			$u_\mathsf{om} \gets o$, where $o \in \symAllIRIs \cup \symAllBNodes$ such that $(u_\textsf{pom},\ttlSmaller{rr:objectMap},o) \in \symRMLGraph_\mathsf{NF}$ \;

			$\omega \gets \extOpAttr{\subjAttr\!}(\omega)$,
				where
			$\extExpr = \textsc{CreateExtExpr}(u_\textsf{sm}, \symRMLGraph_\mathsf{NF}, \baseIRI, \LTB, \symAttrQueryMap)$
			\tcp*[r]{Alg.\ref{alg:rml_termmap_extend_expr}}
			\label{line:rml2algebra:extend_for_subject}

			$\omega \gets \extOpAttr{\predAttr\!}(\omega)$,
				where
			$\extExpr = \textsc{CreateExtExpr}(u_\textsf{pm}, \symRMLGraph_\mathsf{NF}, \baseIRI, \LTB, \symAttrQueryMap)$
			\;
			\label{line:rml2algebra:extend_for_predicate}

			\If{%
				there is a $u_\mathsf{ptm} \in \symAllIRIs \cup \symAllBNodes$ such that $(u_\mathsf{om},\ttlSmaller{rr:parentTriplesMap},u_\mathsf{ptm}) \in \symRMLGraph_\mathsf{NF}$
				\label{line:rml2algebra:join_case_begin}
			}
            {

				$\omega'\! \gets \sourceOp{\dataSource'\!}{\queryExpr'\!}{\symAttrQueryMap'}$\!, where
				$(\dataSource'\!, \queryExpr') = \textsc{SrcAndRootQuery}(u_\mathsf{ptm},\symRMLGraph_\mathsf{NF})$
				\hspace*{25mm} and $\symAttrQueryMap'\! = \textsc{ExtractQueries}(u_\mathsf{ptm},\symRMLGraph_\mathsf{NF})$
				\tcp*{Algorithm~\ref{alg:query_extraction}}
				\label{line:rml2algebra:source2}

				$\symJoinAttrPairs \gets \emptyset$ \;
				\label{line:rml2algebra:init_join_pairs}

				\ForEach{$u_\mathsf{jc} \in \symAllIRIs \cup \symAllBNodes$ for which
				$(u_\mathsf{om},\ttlSmaller{rr:joinCondition},u_\mathsf{jc}) \in \symRMLGraph_\mathsf{NF}$}
                {

					$\attr \gets \symAttrQueryMap^{-1}(\lex)$, where $(u_\mathsf{jc},\ttlSmaller{rr:child},o) \in \symRMLGraph_\mathsf{NF}$ with $o = (\lex,\dt) \in \symAllLiterals$ \;

					$\attr'\! \gets \symAttrQueryMap'^{-1}(\lex)$, where $(u_\mathsf{jc},\ttlSmaller{rr:parent},o) \in \symRMLGraph_\mathsf{NF}$ with $o = (\lex,\dt) \in \symAllLiterals$ \;

					$\symJoinAttrPairs \gets \symJoinAttrPairs \cup \{ (\attr, \attr') \}$ \;
                }
                \label{line:rml2algebra:populate_join_pairs_end}

				$\omega \gets \fctEquiJoinOpBig{\omega}{\omega'}$
				\label{line:rml2algebra:add_equijoin}

				$u_\mathsf{sptm} \gets o$, where $o \in \symAllIRIs \cup \symAllBNodes$ such that $(u_\mathsf{ptm},\ttlSmaller{rr:subjectMap},o) \in \symRMLGraph_\mathsf{NF}$ \;

				$\omega \gets \extOpAttr{\objAttr\!}(\omega)$, where $\extExpr = \textsc{CreateExtExpr}(u_\mathsf{sptm}, \symRMLGraph_\mathsf{NF}, \baseIRI, \LTB, \symAttrQueryMap')$ \;
            }
            \label{line:rml2algebra:join_case_end}
            \Else{
				$\omega \gets \extOpAttr{\objAttr\!}(\omega)$, where $\extExpr = \textsc{CreateExtExpr}(u_\textsf{om}, \symRMLGraph_\mathsf{NF}, \baseIRI, \LTB, \symAttrQueryMap)$ \;
				\label{line:rml2algebra:extend_for_ordinary_object}
            }

			\If{there is a $u_\mathsf{gm} \in \symAllIRIs \cup \symAllBNodes$ such that
            $(u_\textsf{sm},\ttlSmaller{rr:graphMap},u_\mathsf{gm}) \in \symRMLGraph_\mathsf{NF}$ or \hspace*{47.5mm} $(u_\textsf{pom},\ttlSmaller{rr:graphMap},u_\mathsf{gm}) \in \symRMLGraph_\mathsf{NF}$
            \label{line:rml2algebra:extend_for_graph_map_begin}
            }
            {
				$\omega \gets \extOpAttr{\graphAttr\!}(\omega)$, where $\extExpr = \textsc{CreateExtExpr}(u_\mathsf{gm}, \symRMLGraph_\mathsf{NF}, \baseIRI, \LTB, \symAttrQueryMap)$ \;
            }
            \label{line:rml2algebra:extend_for_graph_map_end}
            \Else{
				$\omega \gets \extOpAttr{\graphAttr\!}(\omega)$, where $\extExpr$ is the IRI \ttlSmaller{rr:defaultGraph} \;
				\label{line:rml2algebra:default_graph}
            }

			$\omega_\mathsf{spog} \gets \fctUnionBig{\omega_\mathsf{spog}}{ \fctProjectOp{ \{\subjAttr,\predAttr,\objAttr,\graphAttr\} }{\omega} }$ \;
			\label{line:rml2algebra:union}

        \label{line:rml2algebra:main_loop_end}
        }
        \Return{$\omega_\mathsf{spog}$} \label{line:rml2algebra:return}

\end{algorithm}

\smallskip\noindent\textbf{\itshape Input and Output.}
Since RML mappings are captured as RDF graphs that use
	the RML ontology~\cite{IglesiasMolina2023:RMLOntology}%
, the main input to our algorithm is such an RDF graph%
	~(we assume it uses IRIs of the RML ontology only in the way as intended by that ontology)%
. A second input is an IRI considered as a base for resolving relative IRIs~\cite{Dimou2024:RMLSpec}. As a third input, we assume an injective function~%
	$\LTB \!: \symAllStrings \rightarrow \symAllBNodes$
that maps every string to a unique blank node (which shall become relevant later). The output
is a mapping relation that is obtained by
	applying
operators of our algebra and that can be converted into an RDF dataset as per Definition~\ref{def:generated_dataset}.

\smallskip\noindent\textbf{\itshape Normalization Phase.}
To minimize the
	variations of RML descriptions
%
	to be considered, the algorithm
begins by converting the given RML mapping into a \emph{normal form} in which i)~no shortcut properties
are used, ii)~%
	all referencing object maps have
a join condition, and iii)~every triples map has only a single predicate-object map with a single predicate map, a single object map, and at most one graph map%
.
\enlargethispage{\baselineskip}
Every RML mapping can be converted into this normal form by applying the following sequence of rewriting steps, which we have adopted from Rodríguez-Muro and Rezk~\cite{Mariano2015SPARQL2SQL}, with adaptations specific to RML.
While we present these steps informally,
Appendix~\ref{appdx:queries} captures them formally in terms of update~queries.
None of these steps changes the meaning of the defined~mapping.

\begin{enumerate}
	\item
	\label{item:NormalizationStep:ClassExpansion}
	Every class IRI definition is expanded into a predicate-object map.

	\item
	\label{item:NormalizationStep:ConstantsExpansion}
	All shortcut properties for constant-valued term maps are expanded.

	\item
	\label{item:NormalizationStep:POMDuplication}
	Predicate-object maps with multiple predicate maps or multiple object maps are duplicated to have a single predicate map and a single object map each.

	\item
	\label{item:NormalizationStep:JoinRemoval}
	Referencing object maps without a join condition are replaced by an object map with the subject IRI of the parent triples map as the reference value.

	\item
	\label{item:NormalizationStep:TMDuplication1}
	Every triples map with multiple predicate-object maps is duplicated such that each of the resulting triples maps has a single predicate-object map~only.

	\item
	\label{item:NormalizationStep:TMDuplication2}
	Every triples map with multiple graph maps is duplicated such that each of the resulting triples maps has one of these graph maps.
\end{enumerate}

\noindent\textbf{\itshape Main Loop.}
After normalization, the algorithm iterates over all triples maps defined by the (normalized) RML mapping (lines~\ref{line:rml2algebra:main_loop_begin}--\ref{line:rml2algebra:main_loop_end}). For each of them, it
	combines algebra operators, as described below, to define a mapping relation that, in the end, has the attributes $\subjAttr$, $\predAttr$, $\objAttr$, and $\graphAttr$.
These mapping relations are combined using union (line~\ref{line:rml2algebra:union}), resulting in a mapping relation (line~\ref{line:rml2algebra:return})
	that captures the complete RDF dataset defined by the given RML mapping.

\enlargethispage{\baselineskip}

\smallskip\noindent\textbf{\itshape Translation of the Logical Source.}
As the first step for each triples map, the algorithm creates a source operator
(line~\ref{line:rml2algebra:source}). The first two parameters for this operator---the data source~$\dataSource$ and the query expression $\queryExpr$ for selecting relevant context objects (cf.\ Definition~\ref{def:source_operator})---depend on the logical source defined for the triples map. Since RML is extensible regarding the types of logical sources,
	we have to
abstract the creation of these two parameters by the following function.

\begin{definition}
	\label{def:initSrcAndRootQuery}
	\normalfont
	Given an IRI or blank node $\triplesMapIRI_\mathsf{tm}$ (assumed to denote an RML triples map) and an RDF graph~$\symRMLGraph$ (assumed to describe an RML mapping, including
	$\triplesMapIRI_\mathsf{tm}$),
	$\textsc{SrcAndRootQuery}(\triplesMapIRI_\mathsf{tm},\symRMLGraph)$ is a pair $(\dataSource, \queryExpr)$ of a data source $\dataSource = \dataSourceTuple$ with $\sourceType = \sourceTypeTuple$ and a query expression~$\queryExpr \in \symDataAcc$, as extracted from the description of the logical source of $\triplesMapIRI_\mathsf{tm}$~in~$\symRMLGraph$.
\end{definition}


\vspace{-2mm} 

\begin{example}
	\label{ex:initSrcAndRootQuery}
		Given
	an RDF graph $\symRMLGraph_\mathsf{ex}$ that
		contains
	the
		following RML mapping%
	, $\textsc{SrcAndRootQuery}(\ttl{ex:tm},\symRMLGraph_\mathsf{ex}) = (\dataSource_\mathsf{ex}', \varepsilon)$ where $\dataSource_\mathsf{ex}' = (\sourceTypeX{\mathsf{csv}},\bigDataObject_\mathsf{ex}')$ such that $\sourceTypeX{\mathsf{csv}}$ is the source type for CSV-based data sources (cf.\ Example~\ref{ex:source_type_csv}), $\bigDataObject_\mathsf{ex}'$ is the CSV file mentioned in the second triple, and $\varepsilon$ is the empty string.
	{\scriptsize%
	\begin{verbatim}
ex:tm rml:logicalSource [ rml:source "data.csv";  rml:referenceFormulation ql:CSV ];
      rr:subjectMap [ rr:template "http://example.com/person_{ID}";  rr:termType rr:IRI ];
      rr:predicateObjectMap [ rr:predicate rdfs:label;
                              rr:objectMap [rml:reference "Name"] ].
	\end{verbatim}%
	}
\end{example}

The third parameter for the source operator---the
	partial
function~$\symAttrQueryMap$
	that maps attributes to query expressions
(cf.\ Definition~\ref{def:source_operator})---is populated by
	Algorithm~\ref{alg:query_extraction}, which
extracts the query expressions
	from the term maps of
the given triples map. In particular, the term maps may use query expressions as references (lines~\ref{line:query_extraction:from_references_begin}--\ref{line:query_extraction:from_references_end} of Algorithm~\ref{alg:query_extraction}), in
	string
templates (lines~\ref{line:query_extraction:from_templates_begin}--\ref{line:query_extraction:from_templates_end}%
), and
	as child and parent queries of
join conditions (lines~\ref{line:query_extraction:from_joins_begin}--\ref{line:query_extraction:from_joins_end}%
).
	To focus only on the term maps of the given triples map, the extraction is restricted to the relevant
subgraph of the given RDF graph (line~\ref{line:query_extraction:subgraph} of Algorithm~\ref{alg:query_extraction})%
	, which we define formally~as~follows.

\begin{algorithm}[t]
    \SetArgSty{textnormal}
    \DontPrintSemicolon
	\caption{\textsc{ExtractQueries} - extract query expressions from triples maps.}
	\label{alg:query_extraction}
        \KwData{%
            $\triplesMapIRI_\textsf{tm}$ - IRI or blank node of the triples map
	            to extract queries from \;
			\hspace{10mm}
		    $\symRMLGraph$ - an RDF graph (assumed to contain a \emph{normalized} RML description)
        }

        \KwResult{an injective partial function $\symAttrQueryMap$ that maps attributes to query expressions}

		$\symRMLGraph_\mathsf{tm} \gets$ the $\triplesMapIRI_\mathsf{tm}$-rooted subgraph of $\symRMLGraph$ \tcp*{Definition~\ref{def:rooted_subgraph}}
		\label{line:query_extraction:subgraph}

        $Q \gets \emptyset$ \tcp*{Initially empty set of query expressions}

		\ForAll{$s \in \symAllIRIs \cup \symAllBNodes$ and $o \in \symAllLiterals$ for which there is a triple $(s, \ttlSmaller{rml:reference},o) \in \symRMLGraph_\mathsf{tm}$
		\label{line:query_extraction:from_references_begin}
		}
        {
		    $Q \gets Q \cup \{ \lex \}$, where $o = (\lex,\dt)$ \;
        }
        \label{line:query_extraction:from_references_end}

		\ForAll{$s \in \symAllIRIs \cup \symAllBNodes$ and $o \in \symAllLiterals$ for which there is a triple $(s, \ttlSmaller{rr:template}, o) \in \symRMLGraph_\mathsf{tm}$
		\label{line:query_extraction:from_templates_begin}
		}
        {
			$Q \gets Q \cup Q'\!$, where $o = (\lex,\dt)$ and $Q'$
				is the set of
			all substrings of $\lex$ that \hspace*{16mm} are enclosed by "\{" and "\}" \;
        }
        \label{line:query_extraction:from_templates_end}
		
		$T_\textrm{c} \gets \{ (s,p,o) \in \symRMLGraph_\mathsf{tm} \, | \, p \text{ is the IRI } \ttlSmaller{rr:child} \text{ and } o \in \symAllLiterals \}$ \;
		\label{line:query_extraction:from_joins_begin}

		$T_\textrm{p} \gets \{(s,p,o) \in \symRMLGraph \, | \, p \text{ is the IRI } \ttlSmaller{rr:parent}, o \in \symAllLiterals, \text{ and there is an } s'\! \in \symAllIRIs\cup\symAllBNodes \text{ s.t.}$ \hspace*{26mm} $(s'\!, \ttlSmaller{rr:joinCondition}, s) \!\in\! \symRMLGraph \text{ and } (s'\!, \ttlSmaller{rr:parentTriplesMap}, \triplesMapIRI_\textsf{tm}) \!\in\! \symRMLGraph \}$ \;

        \ForEach{triple $(s,p,o) \in (T_\textrm{c} \cup T_\textrm{p})$}
        {
		    $Q \gets Q \cup \{ \lex \}$, where $o = (\lex,\dt)$ \;
        }
        \label{line:query_extraction:from_joins_end}

		$\symAttrQueryMap \gets$ (initially) empty function from attributes to query expressions \;

		\ForEach{query expression $\queryExpr \in Q$}
		{
			$\symAttrQueryMap \gets \symAttrQueryMap \cup \{ a \rightarrow \queryExpr \}$, where $\attr$ is a fresh attribute from $\symAttrUniverse \setminus \{ \subjAttr, \predAttr, \objAttr, \graphAttr \}$ that \hspace*{24mm} has not been used before in the whole translation process
		}

		\Return{$\symAttrQueryMap$}

\end{algorithm}

\begin{definition}
	\label{def:rooted_subgraph}
	\normalfont
	Let $\symRMLGraph$ be an RDF graph and
		$\triplesMapIRI \in \symAllIRIs \cup \symAllBNodes$.
	The \textbf{$\triplesMapIRI$-rooted subgraph of $\symRMLGraph$}
	is an RDF graph $\symRMLGraph'\! \subseteq \symRMLGraph$
	that is defined recursively as follows:%
\vspace{-2mm} 
	\begin{enumerate}
		\item $\symRMLGraph'$ contains every triple $(s,p,o) \in \symRMLGraph$ for which it holds that $s = \triplesMapIRI$.
		\item $\symRMLGraph'$ contains every triple $(s,p,o) \in \symRMLGraph$ for which there
			is already another
		triple $(s'\!,p'\!,o') \in \symRMLGraph'$ such that $s = o'$ and $p'$ is not the IRI \ttl{rr:parentTriplesMap}.
	\end{enumerate}
\end{definition}

\vspace{-3mm} 

\begin{example}
	\label{ex:ExtractQueries}
	For the RDF graph $\symRMLGraph_\mathsf{ex}$ that contains the RML mapping of
		Example~\ref{ex:initSrcAndRootQuery},
	$\textsc{ExtractQueries}(\ttl{ex:tm},\symRMLGraph_\mathsf{ex})$ returns $\symAttrQueryMap = \{ \attr_1 \rightarrow \textsf{\small ID}, \attr_2 \rightarrow \textsf{\small Name} \}$,
	where $\attr_1$ and $\attr_2$ are arbitrary attributes such that $\attr_1 \neq \attr_2$ and $\attr_1, \attr_2 \notin \{ \subjAttr, \predAttr, \objAttr, \graphAttr \}$.
\end{example}

\enlargethispage{\baselineskip}
\vspace{-1mm} 

\noindent\textbf{\itshape Translation of the Subject Map and the Predicate Map.}
The source operator constructed
in the previous step creates a mapping relation with values obtained via the extracted query expressions (along the lines of Example~\ref{ex:source_operator}).
	This relation can now be extended by using these values
to create RDF terms as defined by the term maps of the triples map. To this end, for the
	subject~(term) map and the predicate~(term)
map, the algorithm adds an extend operator, respectively~(lines~\ref{line:rml2algebra:extend_for_subject}--\ref{line:rml2algebra:extend_for_predicate} in Algorithm~\ref{alg:rml2algebra}). The construction of the extend expressions of these operators
	(see Definitions~\ref{def:ExtendExpressions:Syntax} and~\ref{def:extend_operator})
is captured in Algorithm~\ref{alg:rml_termmap_extend_expr} and uses the following extension functions, which are defined formally in Appendix~\ref{appdx:functions}.
\vspace{-2mm} 
\begin{itemize}
	\item
	$\toIRI$
	converts string literals representing IRIs into these IRIs. 

	\item
	$\toBNode$%
		, parameterized by the aforementioned function $\LTB$ (see the discussion of the input of Algorithm~\ref{alg:rml2algebra}), converts string literals into blank~nodes.

	\item
	$\toLiteral$ is a binary extension function that converts string literals into literals with a datatype IRI that is given as the second argument.

	\item
	$\concat$ is a binary extension function that concatenates two string literals. 
\end{itemize}

\begin{algorithm}[t]
    \SetArgSty{textnormal}
    \DontPrintSemicolon
	\caption{\textsc{CreateExtExpr} -
		creates an extend expression for a term~map.}
	\label{alg:rml_termmap_extend_expr}
        \KwData{%
            $u$ - IRI or blank node of the term map to create the extend expression for \;
            \hspace{10mm}
            $\symRMLGraph$ - an RDF graph (assumed to describe
            	the term map~$u$)\;
            \hspace{10mm}
            $\baseIRI$ - an IRI considered as base IRI \;
            \hspace{10mm}
            $\LTB$ - an injective function
            	that maps
            every string to a unique blank~node \;
            \hspace{10mm}
            $\symAttrQueryMap$ - an injective partial function that maps attributes to query expressions \;
        }
        \KwResult{an extend expression (as per Definition~\ref{def:ExtendExpressions:Syntax})}

		\lIf{%
			$\symRMLGraph$ contains a triple $(u, \ttlSmaller{rr:constant}, o)$
			\label{line:rml_termmap_extend_expr:from_constant_begin}
		}
		{
			\Return{$o$}
		}
		\label{line:rml_termmap_extend_expr:from_constant_end}

		\smallskip

		$\extExpr \gets $ initially empty extend expression \; 

		\If{%
			$\symRMLGraph$ contains a triple $(u, \ttlSmaller{rr:reference}, o)$ such that $o$ is a literal $\literalTuple$
			\label{line:rml_termmap_extend_expr:from_reference_begin}
		}
		{
			$\extExpr \gets a$, where $a$ is the attribute in $\fctDom{\symAttrQueryMap}$ such that $\symAttrQueryMap(a) = \lex$
		}
     \label{line:rml_termmap_extend_expr:from_reference_end}
     
		\ElseIf{
			$\symRMLGraph$ contains a triple $(u, \ttlSmaller{rr:template}, o)$
				such that
			$o$ is a literal $\literalTuple$
			\label{line:rml_termmap_extend_expr:from_template_begin}
		}
		{
			$\substrSeq \gets \fctTempSubStrs{\lex}$ \tcp*[r]{Definition~\ref{def:tempSubStrs}}
			\label{line:rml_termmap_extend_expr:split_template}
			
			$\extExpr \gets (s_1, \ttlSmaller{xsd:string})$, where $s_1$ is the first element in $\substrSeq$ \;\label{line:rml_termmap_extend_expr:first_substring}

			\ForEach
			(\tcp*[f]{Iterate
				based on the order
			in~%
				$\substrSeq$})
			{$s_i$ in
			$\substrSeq$ without the first element
			\label{line:rml_termmap_extend_expr:concatenation_begin}
			}
			{
				\If{$s_i$ is a string that starts with "\{" and ends with "\}"}
				{
					$\extExpr \gets \fctConcatBig{\extExpr}{\symAttrQueryMap^{-1}(\queryExpr)}$, where $\queryExpr$ is
					$s_i$ without
					enclosing "\{" and "\}" \;
				}
				\Else{
					$\extExpr \gets \fctConcatBig{\extExpr}{(s_i, \ttlSmaller{xsd:string})}$ \;
				}
			}
			\label{line:rml_termmap_extend_expr:concatenation_end}

		}
		\label{line:rml_termmap_extend_expr:from_template_end}

		\smallskip

        \lIf{$\symRMLGraph$ contains the triple $(u,\ttlSmaller{rr:termType}, \ttlSmaller{rr:BlankNode})$}{%
            \Return{$\fctToBNode{\extExpr}$}
        }%
        \smallskip
        \lElseIf{$\symRMLGraph$ contains
        $(u, \ttlSmaller{rr:datatype}, o)$
        	with $o \in \symAllIRIs$}
        {%
            \Return{$\fctToLiteral{\extExpr, o}$}%
        }%
        \smallskip
        \lElseIf{$\symRMLGraph$ contains $(u,\ttlSmaller{rr:termType}, \ttlSmaller{rr:Literal})$}
        {%
             \Return{$\fctToLiteral{\extExpr, \ttlSmaller{xsd:string}}$}%
         }%
        \smallskip
        \lElseIf{$\symRMLGraph$ contains $(t,\ttlSmaller{rr:objectMap}, u)$ with $t \in \symAllIRIs \cup \symAllBNodes$
        \textbf{and} $\symRMLGraph$ contains \hspace*{15mm} \phantom{ } \hspace*{8mm} $(u,\ttlSmaller{rml:reference}, o)$ with $o \in \symAllLiterals$}
        {%
             \Return{$\fctToLiteral{\extExpr, \ttlSmaller{xsd:string}}$}
         }%
        \smallskip
        \lElse 
        {%
            \Return{$\fctToIRI{\extExpr}{\baseIRI}$}
        }

\end{algorithm}

The extend expression
	created by Algorithm~\ref{alg:rml_termmap_extend_expr} depends on whether the given
term map is constant-val\-ued
	(line~\ref{line:rml_termmap_extend_expr:from_constant_begin}%
%
), reference-val\-ued (lines~\ref{line:rml_termmap_extend_expr:from_reference_begin}--\ref{line:rml_termmap_extend_expr:from_reference_end}), or template-val\-ued (lines~\ref{line:rml_termmap_extend_expr:from_template_begin}--\ref{line:rml_termmap_extend_expr:from_template_end}). In the latter case,
	\ExtendedVersion{the extend expression needs to be created such that it instantiates the corresponding template string. To this end, this string}%
	\PaperVersion{the corresponding template string}
is first split into a sequence of substrings (line~\ref{line:rml_termmap_extend_expr:split_template}), defined as follows.

\enlargethispage{\baselineskip}
\vspace{-1mm} 

\begin{definition}
	\label{def:tempSubStrs}
	\normalfont
	For every string $\mathit{tmpl} \in \symAllStrings$,
	we write $\tempSubStrs(\mathit{tmpl})$ to denote a sequence~$\substrSeq$ of strings that is constructed by the following
		two 
	steps.%
\vspace{-2mm} 
	\begin{enumerate}
		\item
		Partition $\mathit{tmpl}$ into a sequence~$\substrSeq$ of \emph{query substrings} and \emph{normal substrings} in the order in which these substrings appear in $\mathit{tmpl}$ from left to right,~where
		\begin{itemize}
			\item a \emph{query substring} is a substring
				starting with an opening curly brace (i.e., "\{") and ending with a closing curly brace (i.e., "\}"), and
			\item a \emph{normal substring} is every substring between two query substrings, as well as
				the one before the first and the one
			after the last query substring.
		\end{itemize}

		\item 
		If $\substrSeq$ is the empty sequence, insert the empty string as a new substring into~$\substrSeq$.

%
	\end{enumerate}
\end{definition}

%
The extend expression is then created to concatenate the
	resulting
substrings (lines~\ref{line:rml_termmap_extend_expr:first_substring}--\ref{line:rml_termmap_extend_expr:concatenation_end}), where every query substring "\{$q$\}" is replaced by the attribute~$\attr$ for which
	it holds that
$\symAttrQueryMap(\attr) = q$, because that attribute holds the values that the source operator obtains for the query
	expression~%
$q$.

\begin{example}
	\label{ex:rml_termmap_extend_expr}
	Assume
		$b_\mathsf{sm}$ and $b_\mathsf{om}$ are the blank nodes denoting the subject map and the object map in RDF graph~$\symRMLGraph_\mathsf{ex}$ in
			Example~\ref{ex:initSrcAndRootQuery}, respectively,
		$\baseIRI$ is an arbitrary base IRI,
	and
		$\symAttrQueryMap = \{ \attr_1 \rightarrow \textsf{\small ID},$ $\attr_2 \rightarrow \textsf{\small Name} \}$
	as in Example~\ref{ex:ExtractQueries}.
		Then,
%
	$\textsc{CreateExtExpr}(b_\mathsf{om}, \symRMLGraph_\mathsf{ex}, \baseIRI, \LTB, \symAttrQueryMap)$ returns the extend expression $\fctToLiteral{\attr_2, \ttl{xsd:string}}$. Moreover, $\textsc{CreateExtExpr}(b_\mathsf{sm}, \symRMLGraph_\mathsf{ex}, \baseIRI, \LTB, \symAttrQueryMap)$ returns
		the extend expression
	$
	\toIRI\bigl(
		\concat\bigl(
				\ell,
			\attr_1
		\bigr),
		\baseIRI
	\bigr)
	$
	where $\ell$ is the literal $(\ttl{"http://example.com/person\_"},$ $\ttl{xsd:string})$.
\end{example}

\enlargethispage{\baselineskip}
\vspace{-1mm} 

\smallskip\noindent\textbf{\itshape Translation of the Object Map.}
For object (term) maps, the algorithm also has to consider the case of
	referencing object maps%
~(lines~\ref{line:rml2algebra:join_case_begin}--\ref{line:rml2algebra:join_case_end} in Algorithm~\ref{alg:rml2algebra}), which capture a join with
	RDF terms defined by the subject map
of another triples map. In this case, another source operator needs to be added
	(for this other triples map)
and
	the mapping relation created by that source operator needs
be joined with the mapping relation produced before (line~\ref{line:rml2algebra:add_equijoin}). To this end, an equijoin operator is used, for which the join attributes are determined based on the join condition defined for the referencing object map~(lines~\ref{line:rml2algebra:init_join_pairs}--\ref{line:rml2algebra:populate_join_pairs_end}).
Ordinary object maps are handled
	like
subject
and predicate maps before~(line~\ref{line:rml2algebra:extend_for_ordinary_object}).

\smallskip\noindent\textbf{\itshape Translation of the Graph Map.}
Finally, if the triples map has a graph map, the algorithm adds an extend operator, with attribute~$\graphAttr$, to create IRIs defined by this graph map (lines~\ref{line:rml2algebra:extend_for_graph_map_begin}--\ref{line:rml2algebra:extend_for_graph_map_end}). If the triples map
	has no
graph map, an extend operator that assigns $\graphAttr$ to the constant IRI \ttl{rr:defaultGraph} is added~(line~\ref{line:rml2algebra:default_graph}).

\vspace{-3mm} 

\section{Algebraic Equivalences}%
\label{sec:equivalences}

\vspace{-2mm} 

In addition to providing a basis to formally define the semantics of a mapping language such as RML, our mapping algebra may also be used as a foundation for implementing such languages in mapping tools.
	More precisely,
such tools may use our algebra
	as a form of language
to internally represent the plans that they create for converting data according to a given mapping description, and then they may use algebraic equivalences to rewrite initial mapping plans into more efficient ones%
	~(with the provable guarantee that the rewritten plans produce the same result!)%
. While an extensive study of such optimization opportunities is beyond the scope of this paper, in this section we show a number of such equivalences and describe how they may be relevant for a plan optimizer.

\vspace{-3mm} 

\subsection{Projection Pushing} \label{ssec:equivalences:ProjectionPushing}

\vspace{-2mm} 

The first group of equivalences focuses on pushing
	project operators
deeper into a given plan. This may be useful to reduce the size of intermediate mapping relations in terms of
	values for attributes
that are not used anymore later in the plan.
	Dropping such attributes from intermediate mapping relations as early as possible reduces the memory space required for executing the mapping
		plan.

As a typical example of an algebraic equivalence that an optimizer may use for this purpose, consider the following result, which shows that a project operator over an extend-based subexpression may be pushed
into this subexpression if the set of projection attributes contains both the extend attribute and the attributes mentioned by the corresponding extend expression (if any).%
\footnote{We provide the proof of Proposition~\ref{prop:FullProjectPushing} in Appendix~\ref{appdx:proof}. The following propositions in this section can be shown in the same way.}

\vspace{-1mm} 

\begin{proposition}
	\label{prop:FullProjectPushing}
	\normalfont
	Let $r = \mappingRel$ be a mapping relation,
	$\attr \in \symAttrUniverse$ be an attribute
		that is not in $\symAttrSubSet$ (i.e., $\attr \notin \symAttrSubSet$),
	$\extExpr$ be an extend expression, and
	$\symProjSet \subseteq \symAttrSubSet$ be a non-empty subset of the attributes of $r$.
	If $( \fctAttrs{\extExpr} \cap \symAttrSubSet ) \subseteq \symProjSet$, then it holds that%
\vspace{-2mm} 
	\begin{equation*}
		\fctProjectOpBig{\symProjSet \,\cup\, \{\attr\}\!}{ \fctExtOp{r} }
		=
		\fctExtOpBig{ \fctProjectOp{\symProjSet}{r} }
		.
	\end{equation*}
\end{proposition}

\vspace{-1mm} 

To satisfy the condition for the equivalence in Proposition~\ref{prop:FullProjectPushing} and, thus, to facilitate projection pushing based on this equivalence, in some cases, it may be necessary to first add another projection with additional projection attributes. The following result shows an equivalence that can be used for this purpose.

\vspace{-1mm} 

\begin{proposition}
	\label{prop:ProjectSplitting}
	\normalfont
	Let $r = \mappingRel$ be a mapping relation and
	$\symProjSet \subseteq \symAttrSubSet$ be a non-empty subset of the attributes of $r$.
	For every $\symProjSet'\! \subseteq \symAttrSubSet$, it holds that%
\vspace{-2mm} 
	\begin{equation*}
		\fctProjectOp{\symProjSet}{r}
		=
		\fctProjectOpBig{\symProjSet}{ \fctProjectOp{\symProjSet\cup\symProjSet'\!}{r} }
		.
	\end{equation*}
\end{proposition}

\vspace{-1mm} 

\ExtendedVersion{ 
As a corollary of
	the previous two results,
it is possible to push some projection attributes into an extend-based subexpression even in cases in which the condition for the equivalence in Proposition~\ref{prop:FullProjectPushing} is not directly satisfied.

\begin{corollary}
	\label{corr:PartialProjectPushing}
	\normalfont
	Let $r = \mappingRel$ be a mapping relation,
	$\attr \in \symAttrUniverse$ be an attribute
		such that $\attr \notin \symAttrSubSet$,
	$\extExpr$ be an extend expression, and
	$\symProjSet \subseteq \symAttrSubSet \cup \{\attr\}$.
	It holds that%
\vspace{-2mm} 
	\begin{equation*}
		\fctProjectOpBig{\symProjSet}{ \fctExtOp{r} }
		=
		\fctProjectOpBig{\symProjSet}{ \fctExtOpBig{ \fctProjectOp{\symProjSet'\!}{r} } }
		,
\vspace{-2mm} 
	\end{equation*}
	where $\symProjSet'\! = \bigl(\symProjSet \setminus \{\attr\} \bigr) \cup \bigl(\fctAttrs{\extExpr} \cap \symAttrSubSet \bigr)$.
\end{corollary}

We emphasize that, while the equivalence in Corollary~\ref{corr:PartialProjectPushing} does not eliminate the initial project operator, applying another projection already before the extend operator may still lead to efficiency gains due to reduced memory requirements, in particular if the added projection can be pushed down further.

} 

While further equivalences for projection pushing can be shown (e.g., into a join),
we also want to highlight the existence of cases in which a project operator can be removed completely. For instance, every project operator
	with a source operator as input
may be removed%
	, as shown by the following result%
.

\vspace{-1mm} 

\begin{proposition}
	\label{prop:ProjectionIntoSource}
	\normalfont
	Let $\dataSource = \dataSourceTuple$ be a data source with
		$\sourceType = (\symSetDataset, \symSetContentA\!,$ $\symSetContentB\!, \symDataAcc, \symDataAcc'\!, \eval, \cbeval, \typeCast)$,
		let
	$\queryExpr \in \symDataAcc$,
		let
	$\symAttrQueryMap$ be a partial function
	$\symAttrQueryMap \!: \symAttrUniverse  \rightarrow \symDataAcc'$,~%
	and
	$\symProjSet \subseteq \fctDom{\symAttrQueryMap}$.
	It holds that
\PaperVersion{%
$
		\fctProjectOpBig{\symProjSet}{ \sourceOpDflt }
		=
		\sourceOp{\dataSource}{\queryExpr}{\symAttrQueryMap'}
$,
}%
\ExtendedVersion{
\vspace{-2mm} 
	\begin{equation*}
		\fctProjectOpBig{\symProjSet}{ \sourceOpDflt }
		=
		\sourceOp{\dataSource}{\queryExpr}{\symAttrQueryMap'}
\vspace{-2mm} 
	\end{equation*}
}%
	where $\symAttrQueryMap'$ is the restriction of $\symAttrQueryMap$ to $\symProjSet$; i.e.,
		$\symAttrQueryMap'(\attr) = \symAttrQueryMap(\attr)$ for all $\attr \in \fctDom{\symAttrQueryMap'} = \symProjSet$.
\end{proposition}

\ExtendedVersion{ 
	As a final (and not particularly surprising) example, a
project operator may also be removed completely (anywhere in a
	mapping
plan) if its set of projection attributes coincides with the attributes of the input relation%
.
\begin{proposition}
	\label{prop:ProjectionRemoval}
	\normalfont
	For every mapping relation $r = \mappingRel$, it holds that
	\begin{equation*}
		\fctProjectOp{\symAttrSubSet}{r}
		=
		r
		.
	\end{equation*}
\end{proposition}
} 

\vspace{-5mm} 
\enlargethispage{\baselineskip}

\subsection{Pushing or Pulling Extend Operators} \label{ssec:equivalences:ExtendPushingOrPulling}

\vspace{-1mm} 

\PaperVersion{%
The extended version of this paper shows a second group of equivalences which focus on moving extend operators either on top of a join-based subplan or under the corresponding join operator (i.e., into one of the
inputs of the join)~\cite{ExtendedVersion}.
}
\ExtendedVersion{
For the second group of equivalences that may be useful when optimizing mapping plans, we focus on extend operators that may be placed either on top of a join-based subplan or under the corresponding join operator (i.e., into one of the two inputs of the join).
\ExtendedVersion{%
	This decision may affect the efficiency of the plan. For instance, if the mapping relation~$r_1$ in $\fctEquiJoinOpBig{ \fctExtOp{r_1} }{ r_2 }$ contains many tuples but only very few of them have join partners in $r_2$ and, thus, the result of this join consists of a few tuples only, then it would be more efficient to extend only these few tuples instead of extending all tuples of $r_1$, including all the ones that are essentially filtered out by the join. In other words, it would be more efficient to pull the extend operator out of the join. In contrast, however, if most of the tuples in $r_1$ have multiple join partners in $r_2$ and, thus, the result of the join contains many more tuples than $r_1$, then it would be more efficient to apply the extend operator directly to $r_1$ rather than pulling it out of the join. Hence, depending on the selectivity of the join, a plan optimizer may decide to keep an extend operator under a join, to pull it out, or also to push an extend operator that is currently on top of a join into that join. The following result shows an equivalence that can be used for this purpose (in either direction).
}%
\PaperVersion{%
	Depending on where such an extend operator is placed and how selective the corresponding join is, the extend expression of the extend operator may have to be evaluated for a greater or a smaller number of mapping tuples, making the extend operator a more or less costly element of the overall mapping plan. Therefore, a plan optimizer may decide to pull an extend operator out of a join, or also to push an extend operator that is currently on top of a join into that join. The following result shows an equivalence that provides a formal basis for such a plan rewriting (in either direction).
}

\begin{proposition}
	\label{prop:ExtendAndJoin}
	\normalfont
	Let $r_1 = \mappingRelSpec{1}$ and $r_2 = \mappingRelSpec{2}$ be mapping relations such that $\symAttrSubSet_1 \cap \symAttrSubSet_2 = \emptyset$,
	let $\symJoinAttrPairs \subseteq \symAttrSubSet_{1} \times \symAttrSubSet_{2}$,
	let $\attr \in \symAttrUniverse$ be an attribute such that $\attr \notin \symAttrSubSet_1 \cup \symAttrSubSet_2$, and
	let $\extExpr$ be an extend expression.
	If
\ExtendedVersion{
	$\fctAttrs{\extExpr} \cap \symAttrSubSet_2 = \emptyset$, then it holds that
	\begin{equation*}
		\fctExtOpBig{ \fctEquiJoinOp{r_1}{r_2} }
		=
		\fctEquiJoinOpBig{ \fctExtOp{r_1} }{r_2}
		,
	\end{equation*}
	and if
}
	$\fctAttrs{\extExpr} \cap \symAttrSubSet_1 = \emptyset$, then it holds that%
\vspace{-1mm} 
	\begin{equation*}
		\fctExtOpBig{ \fctEquiJoinOp{r_1}{r_2} }
		=
		\fctEquiJoinOpBig{r_1}{ \fctExtOp{r_2} }
		.
\vspace{-1mm} 
	\end{equation*}
\end{proposition}

The option to push a particular extend operator into a join (as per Proposition~\ref{prop:ExtendAndJoin}) requires this operator to be directly on top of the join in the mapping plan. Similarly, to pull an extend operator out of a join, this operator must be directly under the join. As this may not always be the case, a plan optimizer may try to move a relevant extend operator closer to a join (from either direction). The equivalence of Proposition~\ref{prop:FullProjectPushing} may be used for this purpose. Another
	option
is to swap two subsequent extend operators, as shown by the following result.

\begin{proposition}
	\label{prop:ExtendSwapping}
	\normalfont
	Let $r = \mappingRel$ be a mapping relation,
	$\attr_1,\attr_2 \in \symAttrUniverse$ be two different attributes that are not in $\symAttrSubSet$ (i.e., $\attr_1 \neq \attr_2$ and $\attr_1,\attr_2 \notin \symAttrSubSet$), and
	$\extExpr_1$ and $\extExpr_2$ be two extend expressions.
	If $\attr_1 \notin \fctAttrs{\extExpr_2}$ and $\attr_2 \notin \fctAttrs{\extExpr_1}$, it holds that%
\vspace{-2mm} 
	\begin{equation*}
		\fctExtOpXBig{\attr_1}{\extExpr_1}{ \fctExtOpX{\attr_2}{\extExpr_2}{r} }
		=
		\fctExtOpXBig{\attr_2}{\extExpr_2}{ \fctExtOpX{\attr_1}{\extExpr_1}{r} }
		.
	\end{equation*}
\end{proposition}
} 

\vspace{-3mm} 

\section{Concluding Remarks and Future Work}%
\label{sec:conclude_remark_fw}

\vspace{-2mm} 

The work presented in this paper opens the door for several new directions of research in the domain of
	mappings-based KG construction.
First and foremost, our approach may be
	used
to formalize
other mapping languages besides RML. In this context we emphasize that, while the five types of
operators
	defined
in this paper are sufficient to capture the expressive power of RML,
the algebra may easily be extended with additional types of operators if needed for a particular language or
	use case.
Similarly, the
algebraic equivalences
	shown in this paper
may be complemented with further equivalences%
. A natural next step then is to
	develop
concrete implementation and optimization techniques based on the algebra and the equivalences. Given the language-agnostic nature of the algebra,
	it even becomes trivial for resulting engines to support multiple~languages.


\vspace{-2mm} 
\begin{credits}
	\subsubsection{\ackname}
	%
		The presented work
	was supported by the imec.icon project PACSOI (HBC.2023.0752), which was co-financed by imec and VLAIO and brings together the following partners: FAQIR Foundation, FAQIR Institute, MoveUP, Byteflies, AContrario, and Ghent University~–~IDLab, and by the European Union's Horizon Europe research and innovation programme under grant agreement no.\ 101058682. 

\vspace{-3mm} 
	\subsubsection{\discintname}
	%
	The authors have no competing interests to declare%
	.
\end{credits}

\bibliographystyle{splncs04}
\bibliography{bibliography}

\clearpage
\appendix

\section{SPARQL Update Queries to Create Normalized RML}
\label{appdx:queries}

\vspace{-2mm}

\noindent
	\begin{minipage}[t]{0.44\textwidth}
		{
		\fontsize{6}{6.5}
		\begin{verbatim}
DELETE { ?sm rr:class ?sm_class . }
INSERT { ?tm rr:predicateObjectMap [
                 rr:predicateMap [
                     rr:termType rr:IRI ;
                     rr:constant rdf:type ] ;
                 rr:objectMap [
                     rr:termType rr:IRI ;
                     rr:constant ?sm_class ] ] }
WHERE { ?tm rr:subjectMap ?sm . 
        ?sm rr:class ?sm_class . }
		\end{verbatim}
		}
		\vspace{-3mm}
		\QueryLabel{lst:sparql_class_expand}{Normalization step~\ref{item:NormalizationStep:ClassExpansion} (expand shortcuts for class IRIs%
		).}
	\end{minipage}%
    \hfill%
    \begin{minipage}[t]{0.52\textwidth}
        {
        \fontsize{6}{6.5}
        \begin{verbatim}
DELETE { ?tm rr:predicateObjectMap  ?pom. 
         ?pom rr:predicateMap ?pm ;
              rr:objectMap ?om ;  rr:graphMap ?gm }
INSERT { ?tm rr:predicateObjectMap [
                     rr:predicateMap ?pm ;
                     rr:objectMap ?om ; 
                     rr:graphMap ?gm ] }
WHERE {    ?tm rr:predicateObjectMap ?pom .
           ?pom rr:predicateMap ?pm ;
           ?pom rr:objectMap ?om
 OPTIONAL {?pom rr:graphMap ?gm} } 
        \end{verbatim}
        }
        \vspace{-5mm}
        \QueryLabel{lst:sparql_deduplication}{Normalization step~\ref{item:NormalizationStep:POMDuplication} (duplicate multi-predicate-object maps
            \ into singletons%
        ).}
    \end{minipage}

\vspace{-2mm}
{
\fontsize{6}{6.5}
\begin{verbatim}
DELETE {  ?tm   rr:subject ?sm_constant .   ?pompm rr:predicate ?pm_constant .
          ?termMap rr:graph ?gm_constant .  ?pomom rr:object  ?om_constant .   }
INSERT { ?tm      rr:subjectMap   [ rr:constant ?sm_constant ].
         ?pompm   rr:predicateMap [ rr:constant ?pm_constant ].
         ?pomom   rr:objectMap    [ rr:constant ?om_constant ].
         ?termMap rr:graphMap     [ rr:constant ?gm_constant ]. }
WHERE { { ?tm      rr:subject   ?sm_constant }
  UNION { ?pompm   rr:predicate ?pm_constant }
  UNION { ?pomom   rr:object    ?om_constant }
  UNION { ?termMap rr:graph     ?gm_constant }  }
\end{verbatim}
}
\vspace{-2mm}
\QueryLabel{lst:sparql_shortcut}{Normalization step~\ref{item:NormalizationStep:ConstantsExpansion} (expand shortcuts for constant-valued term maps).}

\vspace{2mm}
\noindent
    \begin{minipage}[t]{0.55\textwidth}
{
\fontsize{6}{6.5}
\begin{verbatim}
DELETE { ?om rr:parentTriplesMap ?ptm }
INSERT { ?om rr:reference  ?ref ;
             rr:template ?template; 
             rr:constant ?const ;
             rr:termType rr:IRI . }
WHERE { ?om rr:parentTriplesMap ?ptm .
        ?ptm rr:subjectMap ?sm . 
  OPTIONAL{ ?sm rr:reference ?ref }
  OPTIONAL{ ?sm rr:template ?template }
  OPTIONAL{ ?sm rr:constant ?const }
  FILTER NOT EXISTS { ?om rr:joinCondition ?jc } } 
\end{verbatim}
}
\vspace{-3mm}
\QueryLabel{lst:sparql_selfjoin}{Normalization step~\ref{item:NormalizationStep:JoinRemoval} (replace referencing object maps
    without join conditions%
).}
    \end{minipage}%
    \hfill%
    \begin{minipage}[t]{0.4\textwidth}
{
\fontsize{6}{6.5}
\begin{verbatim}
DELETE { ?tm rdf:type rr:TriplesMap ;
             rml:logicalSource ?ls ;
             rr:subjectMap ?sm ;
             rr:predicateObjectMap ?pom }
INSERT { [] rml:logicalSource ?ls ;
            rr:subjectMap ?sm ; 
            rr:predicateObjectMap ?pom }
WHERE { ?tm rml:logicalSource ?ls ;
            rr:subjectMap ?sm ;
            rr:predicateObjectMap ?pom }
\end{verbatim}
}
\vspace{-3mm}
\QueryLabel{lst:sparql_tm_singlepom}{Normalization step~\ref{item:NormalizationStep:TMDuplication1} (duplicate triples maps that contain multiple pred.-ob\-ject maps).}
    \end{minipage}

\vspace{2mm}
\noindent
    \begin{minipage}[t]{0.55\textwidth}
{
\fontsize{6}{6.5}
\begin{verbatim}
DELETE { ?tm rr:predicateObjectMap ?pom .
         ?pom rr:graphMap ?pom_gm .       }
INSERT { [] rml:logicalSource ?ls ;
            rr:subjectMap [
                 rr:reference ?ref ;
                 rr:template ?template ;
                 rr:constant ?const ;
                 rr:termType ?ttype ;
                 rr:graphMap ?sm_gm ;
                 rr:graphMap ?pom_gm ] ; 
            rr:predicateObjectMap ?pom .  }
WHERE { ?tm rml:logicalSource ?ls ; 
            rr:subjectMap ?sm ; 
            rr:predicateObjectMap ?pom .
        ?pom rr:graphMap ?pom_gm .
  OPTIONAL { ?sm rr:graphMap ?sm_gm }
  OPTIONAL { ?sm rr:reference ?ref }
  OPTIONAL { ?sm rr:template ?template }
  OPTIONAL { ?sm rr:constant ?const }
  OPTIONAL { ?sm rr:termType ?ttype }     }
\end{verbatim}
}
\vspace{-2mm}
\QueryLabel{lst:sparql_graphmap_norm:1}{Normalization step~\ref{item:NormalizationStep:TMDuplication2}a (duplicate triples maps with subject maps
    where the pred\-i\-cate-ob\-ject maps
contain multiple graph~maps).}
    \end{minipage}%
    \hfill%
    \begin{minipage}[t]{0.4\textwidth}
{
\fontsize{6}{6.5}
\begin{verbatim}
DELETE { ?sm rr:graphMap ?gm1 . }
INSERT { [] rml:logicalSource ?ls ;
            rr:subjectMap [
                 rr:reference ?ref ;
                 rr:template ?template ;
                 rr:constant ?const ;
                 rr:termType ?ttype ;
                 rr:graphMap ?gm1 ] ;
            rr:predicateObjectMap ?pom }
WHERE { ?tm rml:logicalSource ?ls ;
            rr:subjectMap ?sm ;
            rr:predicateObjectMap ?pom .
        ?sm rr:graphMap ?gm1 ;
            rr:graphMap ?gm2
        FILTER ( ?gm1 != ?gm2 )
  OPTIONAL { ?sm rr:reference ?ref }
  OPTIONAL { ?sm rr:template ?template }
  OPTIONAL { ?sm rr:constant ?const }
  OPTIONAL { ?sm rr:termType ?ttype }  }
\end{verbatim}
}
\vspace{-2mm}
\QueryLabel{lst:sparql_graphmap_norm:2}{Normalization step~\ref{item:NormalizationStep:TMDuplication2}b (duplicate subject maps
   that contain multiple graph~maps%
 ).}
    \end{minipage}

\enlargethispage{\baselineskip}
\vspace{-3mm}

\section{Definition of Relevant Extension Functions} \label{appdx:functions}

\vspace{-2mm}

For every $v \in (\symTermUniverse \cup \error)$, $\dt \in (\symTermUniverse \cup \error)$, $v'\! \in (\symTermUniverse \cup \error)$, 
and $\baseIRI \in \symAllIRIs$, we define:
\begin{equation*}
	\fctToBNode{v} =
	\begin{cases}
		\LTB(\lex)
		& \text{if $v$ is a literal $\literalTuple$
s.t.\
%
$\dt = \ttl{xsd:string}$%
        ,}
        \\ 
		\error
		& \text{else.}
	\end{cases}
\end{equation*}
\begin{align*}
    \fctToLiteral{v, \dt} &= 
    \begin{cases}
        (\lex, \dt) &\text{if $v$ is a literal~$(\lex, \dt')$
s.t.\
%
$\dt = \ttl{xsd:string}$%
        }, \\
        \error & \text{else.}
    \end{cases}
\\
	\fctToIRI{v}{\baseIRI} &=
	\begin{cases}
		\mathit{lex}
		& \text{if $v$ is a literal $\literalTuple$ such that $\lex$ is}
		\\[-1mm]
		& \text{a valid IRI and $\dt$ is the IRI \ttl{xsd:string},}
		\\
		\baseIRI \circ \mathit{lex}
		& \text{if $v$ is a literal $\literalTuple$ such that $\lex$ is not}
		\\[-1mm]
		& \text{a valid IRI, but $\baseIRI \circ \mathit{lex}$ is a valid IRI, and}
		\\[-1mm]
		& \text{$\dt$ is the IRI \ttl{xsd:string},}
		\\
		\error
		& \text{else.}
	\end{cases}
\\
    \fctConcat{v}{v'} &= 
    \begin{cases}
        (\lex \circ \lex'\!, \ttl{xsd:string})  &\text{if $v$ and $v'$ are literals $(\lex,\dt)$ and}
		\\[-1mm]
        & \text{$(\lex'\!,\dt')$, respectively,
          such that}
		\\[-1mm]
        &
          \dt = \dt'\! = \ttl{xsd:string},
        \\ 
        \error &\text{else,}
		\\
    \end{cases}
\end{align*}
where $\lex \circ \lex'$ is the string obtained from concatenating the strings $\lex$ and~$\lex'\!$.

\vspace{-2mm}

\section{Proof of Proposition~\ref{prop:FullProjectPushing}}
\label{appdx:proof}

\vspace{-1mm}

		We
	assume that $( \fctAttrs{\extExpr} \cap \symAttrSubSet ) \subseteq \symProjSet$, and
	let $(\symAttrSubSet_1,\symMappingInst_1) = \fctProjectOpBig{\symProjSet \,\cup\, \{\attr\}\!}{ \fctExtOp{r} }$ and $(\symAttrSubSet_2,\symMappingInst_2) = \fctExtOpBig{ \fctProjectOp{\symProjSet}{r} }$.
	To show that $(\symAttrSubSet_1,\symMappingInst_1) = (\symAttrSubSet_2,\symMappingInst_2)$ we first note that, by Definition~\ref{def:projection_operator}, it holds that $\symAttrSubSet_1 = \symProjSet \cup \{\attr\}$, and by Definition~\ref{def:extend_operator}, $\symAttrSubSet_2 = \symProjSet \cup \{\attr\}$. Hence, $\symAttrSubSet_1 = \symAttrSubSet_2$ and, thus, it remains to show that $\symMappingInst_1 = \symMappingInst_2$.

	We first show that $\symMappingInst_1 \subseteq \symMappingInst_2$, for which we consider an arbitrary mapping tuple~$\mappingTuple \in \symMappingInst_1$ and show that $\mappingTuple \in \symMappingInst_2$.
	Given that $\mappingTuple \in \symMappingInst_1$, by Definition~\ref{def:projection_operator}, it holds that i)~$\fctDom{\mappingTuple}=\symProjSet \cup \{\attr\}$ and ii)~there exists a mapping tuple $\mappingTuple'\! \in \fctExtOp{r}$ such that $\mappingTuple(\attr') = \mappingTuple'(\attr')$ for all $\attr'\! \in \symProjSet \cup \{\attr\}$.
	Since $\mappingTuple'\! \in \fctExtOp{r}$, by Definition~\ref{def:extend_operator}, it holds that i)~$\fctDom{\mappingTuple'} = \symAttrSubSet \cup \{\attr\}$ and ii)~there exists a mapping tuple $\mappingTuple''\! \in r$ such that $\fctEval{\extExpr}{\mappingTuple''} = \mappingTuple'(\attr)$ and $\mappingTuple''(\attr') = \mappingTuple'(\attr')$ for all $\attr'\! \in \symAttrSubSet$.
	Now, let $\mappingTuple'''$ be the restriction of $\mappingTuple''$ to $\symProjSet$; i.e., $\mappingTuple'''\! = \mappingTuple''[\symProjSet]$, which, by Definition~\ref{def:projection_operator}, also means that $\fctDom{\mappingTuple'''} = \symProjSet$ and $\mappingTuple'''\! \in \fctProjectOp{\symProjSet}{r}$.
	Then, since $( \fctAttrs{\extExpr} \cap \symAttrSubSet ) \subseteq \symProjSet$ and $\mappingTuple'''(\attr') = \mappingTuple''(\attr')$ for all $\attr' \in \symProjSet$, it holds that $\fctEval{\extExpr}{\mappingTuple'''} = \fctEval{\extExpr}{\mappingTuple''}$.
	As a consequence, there exists a tuple $\mappingTuple^{(4)}\! \in \fctExtOpBig{ \fctProjectOp{\symProjSet}{r} }$ such that i)~$\fctDom{\mappingTuple^{(4)}} = \fctDom{\mappingTuple'''} \cup \{\attr\}$, ii)~$\mappingTuple^{(4)}(\attr) = \mappingTuple'(\attr)$, and iii)~$\mappingTuple^{(4)}(\attr') = \mappingTuple'''(\attr')$ for all $\attr' \in \symProjSet$.
	Since $\fctDom{\mappingTuple'''} \cup \{\attr\} = \symProjSet \cup \{\attr\} = \fctDom{\mappingTuple}$ and $\mappingTuple'(\attr) = \mappingTuple(\attr)$ and $\mappingTuple'''(\attr') = \mappingTuple''(\attr') = \mappingTuple'(\attr') = \mappingTuple(\attr')$ for all $\attr' \in \symProjSet$, we can conclude that $\mappingTuple^{(4)}\! = \mappingTuple$ and, thus, $\mappingTuple \in \fctExtOpBig{ \fctProjectOp{\symProjSet}{r} }$; i.e., $\mappingTuple \in \symMappingInst_2$.

	Now we show that $\symMappingInst_1 \supseteq \symMappingInst_2$, for which we consider a
	tuple~$\mappingTuple \in \symMappingInst_2$ and show that $\mappingTuple \in \symMappingInst_1$.
	Given that $\mappingTuple \in \symMappingInst_2$, by Definitions~\ref{def:extend_operator} and~\ref{def:projection_operator}, it holds that i)~$\fctDom{\mappingTuple}=\{\attr\} \cup \symProjSet$ and ii)~there exists a mapping tuple $\mappingTuple'\! \in \fctProjectOp{\symProjSet}{r}$ such that $\mappingTuple(\attr) = \fctEval{\extExpr}{\mappingTuple'}$ and $\mappingTuple(\attr') = \mappingTuple'(\attr')$ for all $\attr'\! \in \symProjSet$.
	Since $\mappingTuple'\! \in \fctProjectOp{\symProjSet}{r}$, it holds that i)~$\fctDom{\mappingTuple'}=\symProjSet$ and ii)~there exists a mapping tuple $\mappingTuple''\! \in r$ such that $\fctDom{\mappingTuple''} = \symAttrSubSet$ and $\mappingTuple''(\attr') = \mappingTuple'(\attr')$ for all $\attr'\! \in \symProjSet$.
	Now, let $\mappingTuple'''$ be the mapping tuple for which it holds that i)~$\fctDom{\mappingTuple'''} = \symAttrSubSet \cup \{\attr\}$, ii)~$\mappingTuple'''(\attr) = \fctEval{\extExpr}{\mappingTuple''}$, and iii)~$\mappingTuple'''(\attr') = \mappingTuple''(\attr')$ for all $\attr'\! \in \symAttrSubSet$. Hence, by Definition~\ref{def:extend_operator}, $\mappingTuple'''\! \in \fctExtOp{r}$.
	Next, let $\mappingTuple^{(4)}$ be the restriction of $\mappingTuple'''$ to $\symProjSet \cup \{\attr\}$; i.e., $\mappingTuple^{(4)}\! = \mappingTuple'''[\symProjSet \cup \{\attr\}]$, which, by Definition~\ref{def:projection_operator}, means that $\mappingTuple^{(4)}\! \in \fctProjectOpBig{\symProjSet \,\cup\, \{\attr\}\!}{ \fctExtOp{r} }$.
	Putting everything together, we have that i)~$\fctDom{\mappingTuple^{(4)}} = \symProjSet \cup \{\attr\} = \fctDom{\mappingTuple}$, ii)~$\mappingTuple^{(4)}(\attr) = \mappingTuple'''(\attr) = \fctEval{\extExpr}{\mappingTuple''} = \fctEval{\extExpr}{\mappingTuple'} = \mappingTuple(\attr)$, and iii)~$\mappingTuple^{(4)}(\attr') = \mappingTuple'''(\attr') = \mappingTuple''(\attr') = \mappingTuple(\attr)$ for all $\attr'\! \in \symProjSet$. As a consequence, $\mappingTuple^{(4)} = \mappingTuple$ and, thus, $\mappingTuple \in \fctProjectOpBig{\symProjSet \,\cup\, \{\attr\}\!}{ \fctExtOp{r} }$; i.e.,~$\mappingTuple \in \symMappingInst_1$.

\end{document}